\documentclass{article}

\usepackage[english]{babel}

\usepackage[a4paper,top=2cm,bottom=2cm,left=3cm,right=3cm,marginparwidth=1.75cm]{geometry}

\usepackage{placeins}
\usepackage{amssymb}
\usepackage{fvextra}
\usepackage{fancyvrb}
\usepackage{siunitx}
\usepackage[T1]{fontenc}
\usepackage{libertinus}
\PassOptionsToPackage{hyphens}{url}\usepackage{hyperref}
\usepackage[utf8]{inputenc}
\usepackage{amsmath} 
\usepackage{cleveref}
\usepackage{graphicx}
\usepackage{subcaption}
\usepackage{float}
\usepackage[ruled,vlined,linesnumbered]{algorithm2e}
\usepackage{amsmath, amssymb}

\usepackage{csquotes}
\usepackage{booktabs}
\usepackage{longtable}
\usepackage{adjustbox}
\usepackage{array}
\usepackage{url}
\usepackage{titlesec}
\usepackage{authblk}
\usepackage{xcolor} % Load the xcolor package for color options
\usepackage{tikz}
\usetikzlibrary{arrows.meta,positioning,shapes.geometric,shapes.misc}

\titleformat{\subsection}
  {\mdseries\itshape\large} % Medium series, italic shape, and large font size
  {\thesubsection}{1em}{} % Numbering, spacing, and the section title itself

\usepackage[
  backend=biber,
  style=authoryear,
  citestyle=authoryear,
  maxcitenames=3,
  maxbibnames=99,
  giveninits=true,
  uniquename=init,
  uniquelist=false,
  doi=false,
  isbn=false,
  url=true
]{biblatex}

\crefformat{figure}{#2Figure~#1#3}
\Crefformat{figure}{#2Figure~#1#3}
\crefformat{table}{#2Table~#1#3}
\Crefformat{table}{#2Table~#1#3}
\crefformat{section}{#2Section~#1#3}
\Crefformat{section}{#2Section~#1#3}

\author[1]{Sara Zoccheddu\thanks{Corresponding author. Email: sara.zoccheddu@uzh.ch}}
\author[1,2]{Shah Rukh Qasim}
\author[2]{Patrick Owen}
\author[1,2]{Nicola Serra}

\affil[1]{Mathematical Modeling and Machine Learning Department, University of Zurich, Zurich, Switzerland}
\affil[2]{Physik-Institut, University of Zurich, Zurich, Switzerland}

\title{Large Language Models for Physics Instrument Design}

\begin{document}
\maketitle

\begin{abstract}
We study the use of large language models (LLMs) for physics instrument design and compare their performance to reinforcement learning (RL). Using only prompting, LLMs are given task constraints and summaries of prior high-scoring designs and propose complete detector configurations, which we evaluate with the same simulators and reward functions used in RL-based optimization. Although RL yields stronger final designs, we find that modern LLMs consistently generate valid, resource-aware, and physically meaningful configurations that draw on broad pretrained knowledge of detector design principles and particle--matter interactions, despite having no task-specific training. Based on this result, as a first step toward hybrid design workflows, we explore pairing the LLMs with a dedicated trust region optimizer, serving as a precursor to future pipelines in which LLMs propose and structure design hypotheses while RL performs reward-driven optimization. Based on these experiments, we argue that LLMs are well suited as meta-planners: they can design and orchestrate RL-based optimization studies, define search strategies, and coordinate multiple interacting components within a unified workflow. In doing so, they point toward automated, closed-loop instrument design in which much of the human effort required to structure and supervise optimization can be reduced.
\end{abstract}
% \linenumbers

\section{Introduction}
\label{sec:introduction}
Modern physics instruments are among the most complex systems ever designed, requiring sustained innovation long before any component is constructed or installed. The Large Hadron Collider (LHC) at CERN exemplifies this complexity: its realization followed decades of conceptual studies and more than ten years of intensive design and R\&D, with overall costs amounting to several billion Swiss francs. The project demanded the coordinated design of thousands of highly specialized components, many of which required the development of entirely new technologies and design methodologies to meet unprecedented performance and reliability requirements. Looking ahead, proposed next-generation facilities such as the Future Circular Collider (FCC) are expected to require substantially greater resources, with projected costs in the tens of billions and design phases extending over multiple decades. The scale and ambition of these projects imply an extraordinary level of R\&D effort, underscoring the growing importance of systematic, efficient, and well-founded approaches to the design of physics instruments.

The importance of systematic design extends well beyond physics, with similar challenges encountered across a wide range of scientific and engineering disciplines. In fields such as materials science, mechanical and aerospace engineering, and integrated circuit design, early-stage design decisions critically shape achievable performance, cost, and development timelines. In recent years, these communities have increasingly adopted artificial intelligence to support and partially automate the design process. Data-driven surrogate models, Bayesian optimization, and deep generative models, including variational autoencoders and generative adversarial networks, are now routinely used to explore high-dimensional design spaces, perform rapid design-space exploration, and generate candidate solutions under complex constraints~\parencite{Tian2025,Schmidt2019,Regenwetter2022}.

Recent work has increasingly explored RL as a means of addressing design problems defined by sequential and path-dependent decision processes. Unlike supervised or surrogate-model approaches, which typically assume a fixed design parameterization and objective, RL formulates design as a sequential decision process in which an agent incrementally constructs or modifies a design while receiving feedback through a reward signal. This formulation naturally accommodates complex constraints, long-horizon dependencies, and discrete design choices that are difficult to address with conventional optimization techniques. Notable successes include the application of RL to molecular generation, where agents learn to construct chemically valid and functionally optimized molecules through sequential graph or string manipulations~\parencite{Olivecrona2017,popova2018deep}, as well as large-scale integrated circuit layout, where RL has been shown to produce chip placements that rival or surpass human expert designs while dramatically reducing design time~\parencite{mirhoseini2021graph}.

%The first concrete application of machine learning to the design of a particle physics instrument was the optimization of the Muon Shield for the SHiP experiment~\parencite{Akmete2017}. The goal of the Muon Shield is to suppress the muon flux emerging from a high-intensity proton beam dump by approximately six orders of magnitude using a sequence of tapered magnetic elements. The shield consists of multiple magnets, each described by several geometric parameters, resulting in a high-dimensional and strongly constrained optimization problem with highly non-linear dependencies. Initial design studies employed Bayesian Optimization to efficiently explore this complex design space~\parencite{baranov2017optimising}.
The first concrete application of machine learning to instrument design in particle physics concerned the optimization of the SHiP Muon Shield, formulated as a high-dimensional, strongly constrained problem with expensive and highly non-linear objective evaluations~\parencite{Akmete2017}. Bayesian Optimization was shown to efficiently explore this complex design space and identify high-performing solutions with relatively few evaluations~\parencite{baranov2017optimising}. Another first work utilizing Bayesian Optimization includes the optimization of the ring-imaging Cherenkov detectors at the proposed Electron–Ion Collider~\parencite{Cisbani2020}. Subsequent work introduced a gradient-based approach in which a generative adversarial network was trained as a differentiable surrogate~\parencite{shirobokov2020black} for the Geant4 simulation~\parencite{geant4_cite}, enabling end-to-end backpropagation through the design pipeline and significantly accelerating design iterations. This sequence of developments established a foundational paradigm for AI-assisted instrument design in particle physics. Building on this work, similar machine-learning-based optimization strategies have since been applied to other detector design problems, muon tomography systems~\parencite{Strong2024}, and calorimeter optimization studies~\parencite{Schmidt2025}. A comprehensive overview of these efforts, with a focus on differentiable programming approaches, is provided in~\textcite{dorigo2023toward}, which documents the broader methodology developed within the MODE collaboration~\parencite{mode_collaboration}.

As in other scientific and engineering disciplines, there has also been growing interest in exploring RL for particle physics instrument design, motivated by its ability to address structural and sequential design decisions that go beyond the optimization of a fixed set of parameters. Many of the most impactful improvements in particle physics detectors have historically resulted from high-level layout choices, such as the removal or relocation of entire tracking stations in LHCb \parencite{reoptimized_lhcb} or the introduction of downstream veto calorimetry in NA62 \parencite{na62_status}. These types of decisions define combinatorial, mixed discrete–continuous design spaces that are difficult to explore with conventional optimization techniques, including Bayesian or differentiable approaches, which typically assume a predefined instrument structure. RL offers a natural framework for such problems by formulating instrument design as a sequential decision process in which an agent incrementally constructs a detector layout while receiving physics-based feedback. Building on this perspective, recent work has demonstrated the application of RL to particle physics instrument design through a set of concrete benchmark studies \parencite{qasim2024physics-instrument-rl}. In this work, RL is used to construct detector layouts sequentially, allowing both discrete structural choices and continuous design parameters to be optimized simultaneously. Two representative case studies are considered: the longitudinal segmentation of a calorimeter and the joint optimization of the longitudinal placement and transverse segmentation of tracking layers in a magnetic spectrometer. In both examples, the detector configuration is not fixed in advance, and the agent learns layout strategies directly from physics-based performance feedback, illustrating how RL can be applied to flexible, structure-aware instrument design problems.

Despite these advances, RL does not eliminate the need for human intervention in the design process. The structure of the design problem itself, such as the choice of action space, reward definition, constraints, and optimization strategy, must still be specified manually. As detector design problems scale to include multiple subsystems, heterogeneous constraints, and long-horizon dependencies, this manual orchestration increasingly dominates the overall effort. In practice, a substantial fraction of the design work lies not in optimization per se, but in reasoning about how to structure and coordinate the optimization process. LLMs are neural networks based on the transformer architecture~\parencite{attention} with attention mechanisms to enable scalable sequence modeling. While transformer-based models were initially developed for natural language processing tasks such as translation and text generation, their impact expanded dramatically with the emergence of very large models trained on web-scale corpora. This transition entered the public and scientific mainstream with the release of ChatGPT in late 2022~\parencite{openai_chatgpt_2022}, which demonstrated that sufficiently large pretrained models could perform a wide range of tasks in a zero-shot or few-shot setting using natural-language prompts alone. Subsequent work showed that these models exhibit nontrivial reasoning capabilities that are not explicitly trained but instead emerge with scale. In particular, chain-of-thought prompting was shown to elicit multi-step reasoning behavior on arithmetic, symbolic, and logical tasks~\parencite{reasoning2022_wei}, while even simple zero-shot prompts such as “let’s think step by step” were found to substantially improve reasoning performance across diverse benchmarks~\parencite{zero_shot_llms_2022}. More broadly, systematic studies of scaling behavior have identified a class of emergent abilities—capabilities that appear abruptly beyond certain model sizes and are not predictable from smaller models—including reasoning, abstraction, and planning~\parencite{wei_emergent_abilities_llms_2022}. These developments have motivated growing interest in applying LLMs beyond traditional language-centric tasks, particularly in domains that require structured decision making under constraints. Recent work has explored the use of LLMs in engineering and design contexts, highlighting their potential to support conceptual design, requirement reasoning, and cross-domain synthesis~\parencite{Chiarello2024,Gpfert2024}. Concrete applications have begun to appear across multiple fields, including multimodal CAD generation~\parencite{Li2024}, electronic design automation and VLSI optimization~\parencite{Lu2025}, and molecular and drug design~\parencite{Zheng2025}.

In contrast to RL--based design, this work investigates a planning-oriented approach in which candidate detector layouts are proposed by an LLM, while the evaluation of each design remains identical to that of \textcite{qasim2024physics-instrument-rl}. Design proposals are generated under the same constraints and performance objectives and are evaluated using the same physics simulations and reward definitions. In addition to direct proposal-and-evaluate search, we include a limited hybrid variant in which LLM-proposed designs are passed to a dedicated trust region optimizer for refinement. This is intended as a preliminary demonstration of how LLM proposal generation can be integrated with RL-based optimization. The approach is benchmarked on the same calorimeter and spectrometer design problems as the RL baseline, ensuring that any observed differences arise from the proposal and exploration strategy rather than from changes in the underlying detector models or performance metrics. The methodology is described in Section~\ref{sec:methodology}, and the empirical results are presented in Section~\ref{sec:results}. The broader implications of this hybrid perspective are discussed in Section~\ref{sec:outlook}, with conclusions in Section~\ref{sec:conclusion}.

\section{Design Problems}
\label{sec:design_problems}

We consider the same two instrument-design benchmark problems introduced in
\textcite{qasim2024physics-instrument-rl}. In both cases, the detector models,
simulation procedures, reconstruction algorithms, and reward definitions are
identical to those used in the RL study. The present
work therefore does not introduce new physics models or new optimization
objectives, but instead reuses the same design problems as controlled testbeds
for studying alternative design-generation strategies.

The benchmarks are intentionally simplified and are not meant to represent
fully realistic detector simulations. Rather, they define structured, mixed
discrete--continuous design spaces under global resource constraints, evaluated
using physics-motivated performance metrics. This makes them well suited for
methodological comparisons between RL and LLM-based design
approaches. Full technical details of the simulation and reconstruction
implementations are provided in \textcite{qasim2024physics-instrument-rl}; here
we summarize the design variables, reconstruction logic, and scoring functions
required to define the optimization problems.

\subsection{Calorimeter design benchmark}
\label{sec:method_calo}

The first benchmark concerns the longitudinal segmentation of a sampling
calorimeter. A candidate design is defined by a sequence of active sensor layers
placed along the longitudinal coordinate $z$ within a fixed absorber volume.
Each layer is assigned one of a small number of discrete sensor types, which
differ in effective thickness and cost. A global budget constraint limits the
total sensor cost, thereby constraining both the number of layers and their
composition.

For a given design, detector performance is evaluated by resampling
pre-simulated energy-deposit data according to the proposed sensor placement.
This approach ensures that performance depends on the longitudinal layout of
the active material while keeping the evaluation computationally efficient.
After a fixed calibration step, the calorimeter response is evaluated for
electromagnetic and hadronic particles at representative energies.

Performance is quantified using the mean-corrected energy resolution,
\begin{equation}
\Sigma = \frac{\sigma(E_{\mathrm{pred}} / E_{\mathrm{true}})}{\mu(E_{\mathrm{pred}} / E_{\mathrm{true}})}~,
\end{equation}
where $\sigma$ and $\mu$ denote the standard deviation and mean of the response
distribution, respectively.

The scalar design score $S$ is constructed from the four resolution values using
thresholded linear penalties,
\begin{equation}
S \cdot 10 =
-\max(0, \Sigma_{\mathrm{em50}} - 8)
-\max(0, \Sigma_{\mathrm{em100}} - 5)
-\max(0, \Sigma_{\mathrm{had50}} - 25)
-\max(0, \Sigma_{\mathrm{had100}} - 18)~,
\end{equation}
where $\Sigma_{\mathrm{em}}$ and $\Sigma_{\mathrm{had}}$ denote electromagnetic
and hadronic resolutions in percent, and the numerical thresholds correspond to
fixed reference targets. The score is capped for unphysically large resolutions,
and higher values of $S$ correspond to better-performing designs.

\subsection{Spectrometer design benchmark}
\label{sec:method_spectro}

The second benchmark concerns the design of a magnetic spectrometer through the
placement and segmentation of tracking stations. The setup consists of a fixed
magnetic region located at the center of the apparatus, with charged particles
traversing an upstream region (Region~A), the magnet, and a downstream region
(Region~C). A candidate design specifies the longitudinal locations of tracking
stations (continuous variables) together with the transverse granularity of each
station (discrete variables), subject to a global sensor-budget constraint.

Charged-particle trajectories are generated using a simplified model that
includes magnetic deflection and material scattering.

Momentum reconstruction is performed by fitting straight-line track segments
independently in Regions~A and~C, followed by estimating the momentum from the
measured deflection across the magnetic field. Track reconstruction relies on a
pattern-recognition procedure combined with a Kalman-filter–based line fit.
Starting from an initial seed hit, candidate track segments are constructed by
associating hits across successive tracking stations. At each step, the most
compatible hit in the next station is selected, and the track parameters are
updated using a Kalman filter that accounts for measurement uncertainty arising
from the finite sensor granularity as well as stochastic deviations induced by
material scattering. This iterative procedure yields optimal linear estimates of
the upstream and downstream track segments, which are then combined to infer the
particle momentum.

Design performance is summarized by two metrics: momentum resolution and a
tracking efficiency defined as the fraction of particles satisfying
\begin{equation}
0.5 < \frac{p_{\mathrm{measured}}}{p_{\mathrm{true}}} < 2.0~.
\end{equation}
These quantities are evaluated at representative particle energies of $10$ and
$100~\mathrm{GeV}$.

The scalar design score is constructed as
\begin{equation}
S = S_{10} + S_{100}~,
\end{equation}
where the energy-dependent contributions are given by
\begin{equation}
S_{10} =
-3 \cdot \Big(
95 - \min(\mathrm{eff}_{10}, 95)
+ \max(\mathrm{res}_{10}, 3) - 3
\Big)~,
\end{equation}
\begin{equation}
S_{100} =
-3 \cdot \Big(
95 - \min(\mathrm{eff}_{100}, 95)
+ \max(\mathrm{res}_{100}, 8) - 8
\Big)~.
\end{equation}
Here, $\mathrm{eff}$ denotes the tracking efficiency in percent and
$\mathrm{res}$ the momentum resolution in percent. Efficiency penalties vanish
once the efficiency exceeds $95\%$, while resolution penalties decrease linearly
as the design approaches the reference targets of $3\%$ at $10~\mathrm{GeV}$ and
$8\%$ at $100~\mathrm{GeV}$.

Designs that fail to place a minimum number of tracking stations in both Regions
A and C incur an additional penalty, ensuring that only physically
reconstructable layouts achieve competitive scores. 
%All simulation, reconstruction, and scoring details are otherwise identical to those used in \textcite{qasim2024physics-instrument-rl}.

\section{Methodology: Large Language Models for Design}
\label{sec:methodology}

We use the LLMs as a proposal generator for detector configurations. The LLM does not receive gradients, does not interact with the simulator directly, and is not fine-tuned on these tasks. Instead, it is prompted with (i) an explicit description of the design constraints and objectives and (ii) a curated record of previously evaluated designs. The simulator, reconstruction, and reward definitions are identical to those used in the RL baselines; only the mechanism for proposing candidate designs differs.

\paragraph{Information provided to the LLM.}
At each iteration, the model receives two classes of information.

\begin{enumerate}
    \item \textbf{Specification of the design problem.}
    The prompt enumerates hard constraints (units, bounds, discrete design choices, non-overlap/region requirements, and a global budget), together with the optimization targets expressed in the same metrics used by the simulator-based evaluator. The model is instructed to return only a JSON object encoding a complete design:
    \texttt{\{"z":[...],"t":[...]\}} for the calorimeter and
    \texttt{\{"z":[...],"g":[...]\}} for the spectrometer. The full prompt templates are provided in Appendix~\ref{app:prompts}.

    \item \textbf{Previously evaluated designs (memory).}
    The model is also given summaries of the best designs found so far, including their reward and key aggregate properties (e.g., number of layers/stations, total budget usage, and the best metric values achieved). This memory acts as a compact dataset of feasible, simulator-verified solutions that the LLM can imitate, modify, or extrapolate from.
\end{enumerate}

\paragraph{Token-aware context management.} As prior information grows, we keep the prompt within a configured context window $L$  by reserving an overflow margin $R$ and a reply headroom floor $F$. We then greedily append whole design blocks (best reward first) until the estimated prompt length reaches $(L - R - F)$, never truncating a block. Token counts are estimated with a standard tokenizer.

\paragraph{Projection to the feasible set.}
The raw JSON proposed by the LLM is parsed and projected onto the feasible set before evaluation. For the calorimeter, we sort by position, snap layer thicknesses to the allowed discrete set, remove overlaps and out-of-range layers, and enforce the total thickness budget by truncation. For the spectrometer, we sort by position, snap granularities to the allowed discrete set, enforce the longitudinal bounds and the global pixel budget, and ensure required coverage (minimum station counts upstream and downstream of the magnet). This projection step is critical in practice: LLM outputs can be syntactically valid yet violate hard geometric or resource constraints, and without a deterministic cleaning stage the optimization loop would waste most evaluations on invalid designs. By guaranteeing feasibility, the projection makes the search stable, comparable across iterations, and robust to occasional malformed proposals.

\paragraph{Scoring, acceptance, and logging.}
The projected design is evaluated by the same simulator and reward function used in the RL environment. If the obtained reward strictly improves the best value seen so far for that subsystem, the design is marked accepted and added to the set of designs eligible for future context packing. All attempts (accepted or not) are recorded for auditability and reproducibility, including the raw model output, the projected configuration, and the resulting metrics and reward. This yields an iterative search procedure where the LLM proposes designs conditioned on a growing set of validated exemplars, while feasibility and performance are enforced externally by projection and simulation.

\paragraph{Local trust-region refinement.}
As a proof of concept for hybrid optimization workflows, we additionally study a variant in which each feasible design proposed by the LLM is used as the starting point for a dedicated local optimizer. After projection to the feasible set, we optionally apply a derivative-free trust-region (TR) optimizer, BOBYQA \parencite{powell2009bobyqa}, to refine the design using the same black-box simulator-based reward function as in the direct proposal-and-evaluate loop.

\noindent In this work, TR refinement is deliberately restricted to the longitudinal placement variables $z$ only. All other design choices are held fixed, including discrete selections (sensor types or station granularities) and any additional parameters. For the calorimeter benchmark, TR refinement therefore adjusts only the longitudinal positions of the active layers while keeping the sensor-type assignments unchanged. For the spectrometer benchmark, TR refinement optimizes only the tracking-station $z$-positions, while leaving the transverse granularity vector fixed. All candidate updates proposed by the TR optimizer are re-projected to enforce hard bounds, non-overlap conditions and region-coverage requirements before evaluation. %and global budget constraints

\noindent For each valid design proposed by an LLM, we run $100$ consecutive iterations of the TR optimizer, and we retain the best design encountered over the refinement trajectory. This hybrid LLM+TR procedure has been evaluated for the both the calorimeter and the spectrometer benchmark for GPT-OSS-20B, GPT-OSS-120B, GPT-5, and Gemini~2.5~Pro. Pseudocode for the full procedure is given in Appendix~\ref{app:algorithm} (Algorithm~\ref{alg:llm_search}).

\section{Empirical Results}
\label{sec:results}
We evaluate the design-generation procedure of
Section~\ref{sec:methodology} on the two benchmark environments introduced in
Section~\ref{sec:design_problems}. Throughout, the simulator, reconstruction,
and reward definitions are unchanged relative to the RL baselines of
\textcite{qasim2024physics-instrument-rl}; the comparison therefore isolates the
effect of the design-generation and search strategy.

Our results are organized in two stages. We first report the behavior of
standalone LLMs used purely as proposal generators in the proposal--projection--evaluation
loop. This isolates what can be obtained from pretrained model knowledge and
prompted reasoning under explicit constraints, without any task-specific
training or reward-driven credit assignment. We then report results for the
hybrid pipeline in which the TR refinement stage introduced in
Section~\ref{sec:methodology} is applied downstream of the LLM proposals. This
second stage probes how much additional performance can be recovered by adding a
dedicated reward-driven optimizer on top of the LLM prior, and serves as a
minimal demonstration of the modular pattern in which more powerful optimization
components can be placed after an LLM proposal step.

We consider four LLMs spanning both open(-weights) and proprietary systems:
GPT-OSS-20B, GPT-OSS-120B, GPT-5, and Gemini~2.5~Pro. For each model and
each benchmark, we run the same iterative search loop for a fixed number of
proposal iterations, using identical feasibility-projection rules and an
identical acceptance criterion (a design is accepted only if it improves the
best reward seen so far). When TR refinement is used, it is applied
after projection using the same simulator-based reward function, with the
refinement restricted to the longitudinal placement variables as described in
Section~\ref{sec:methodology}.

Performance is assessed primarily through the best-so-far scalar reward as a
function of iteration number. We report the corresponding evolution of the
underlying physics metrics that enter the reward, and we summarize the
best-performing designs achieved within the evaluation budget. Sections
\ref{sec:results_calo} and \ref{sec:results_spectro} present these results for
the calorimeter and spectrometer benchmarks, respectively, enabling direct
comparisons between baseline designs, RL, standalone LLM proposal search, and
LLM+TR refinement.

\subsection{Calorimeter}
\label{sec:results_calo}

In the calorimeter benchmark, we evaluate the LLM-based proposal-and-evaluate procedure on the longitudinal segmentation task introduced in Section~\ref{sec:method_calo}. The same fast calorimeter simulator and physics-driven score definition are used as for the RL baseline, enabling a direct comparison between LLM-guided and RL-based design strategies.

\begin{figure}[ht]
    \centering
    
    % Top-left
    \begin{subfigure}{0.48\textwidth}
        \centering
        \includegraphics[width=\textwidth]{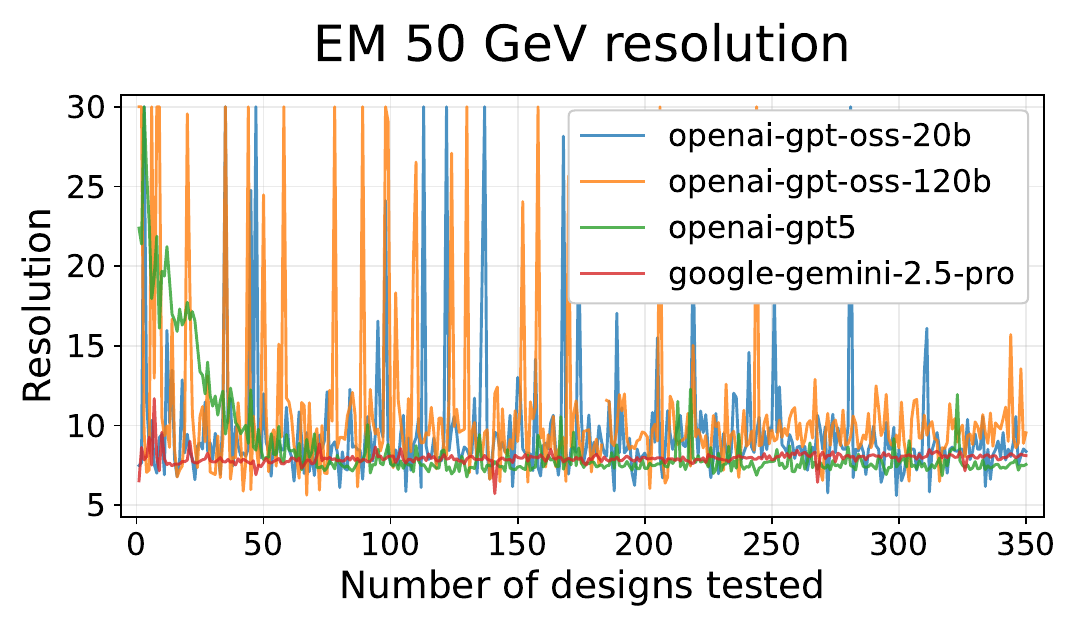} % or .png, .jpg
        % \caption{Caption 1}
    \end{subfigure}
    \hfill
    % Top-right
    \begin{subfigure}{0.48\textwidth}
        \centering
        \includegraphics[width=\textwidth]{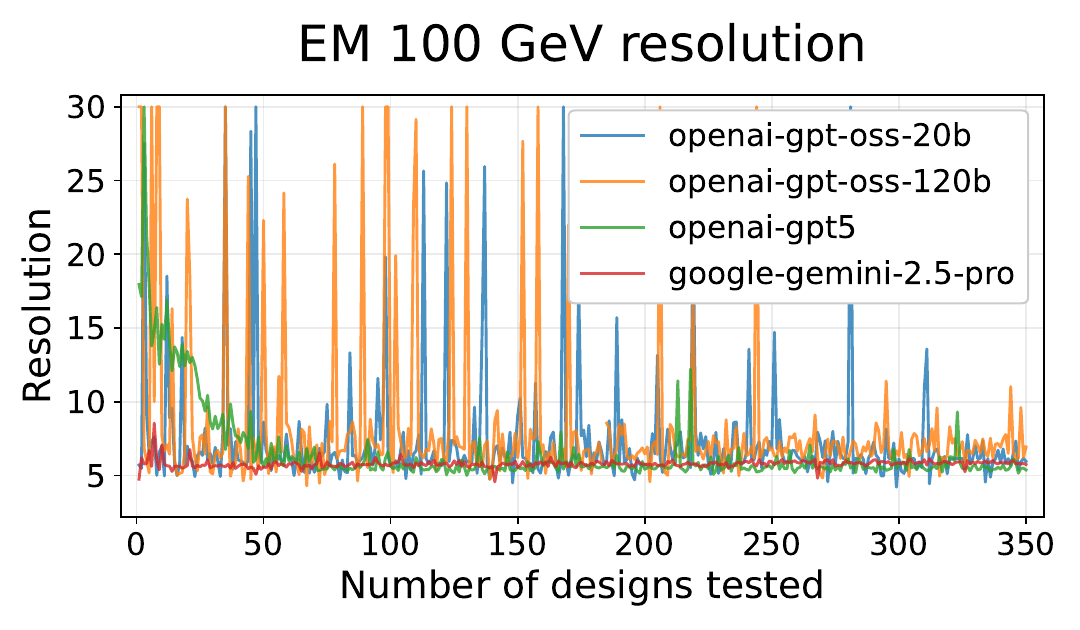}
        % \caption{Caption 2}
    \end{subfigure}

    \vspace{0.5em}

    % Bottom-left
    \begin{subfigure}{0.48\textwidth}
        \centering
        \includegraphics[width=\textwidth]{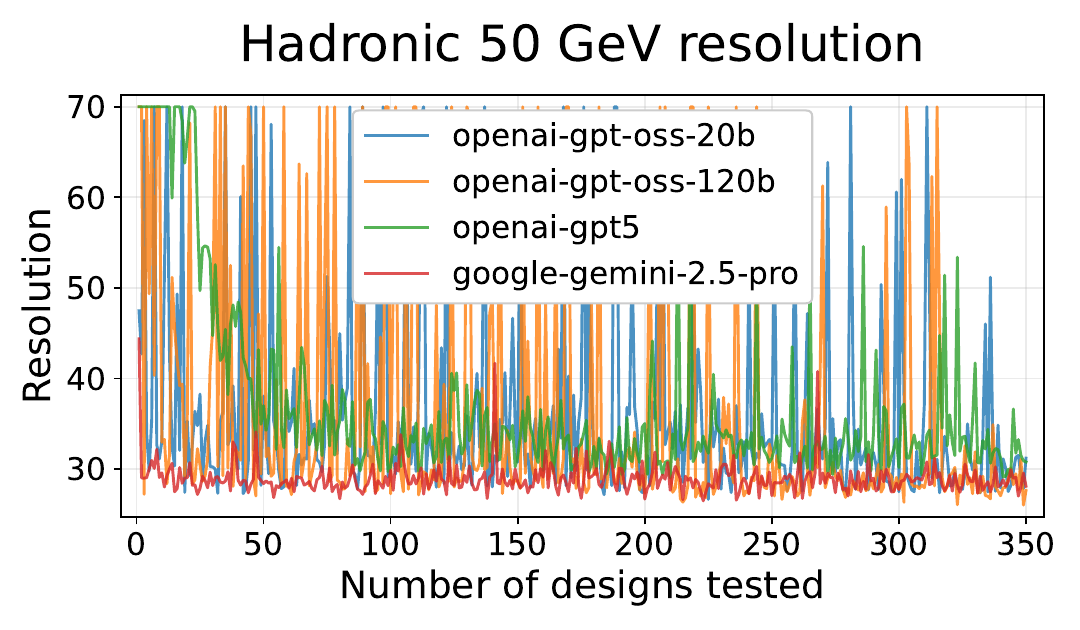}
        % \caption{Caption 3}
    \end{subfigure}
    \hfill
    % Bottom-right
    \begin{subfigure}{0.48\textwidth}
        \centering
        \includegraphics[width=\textwidth]{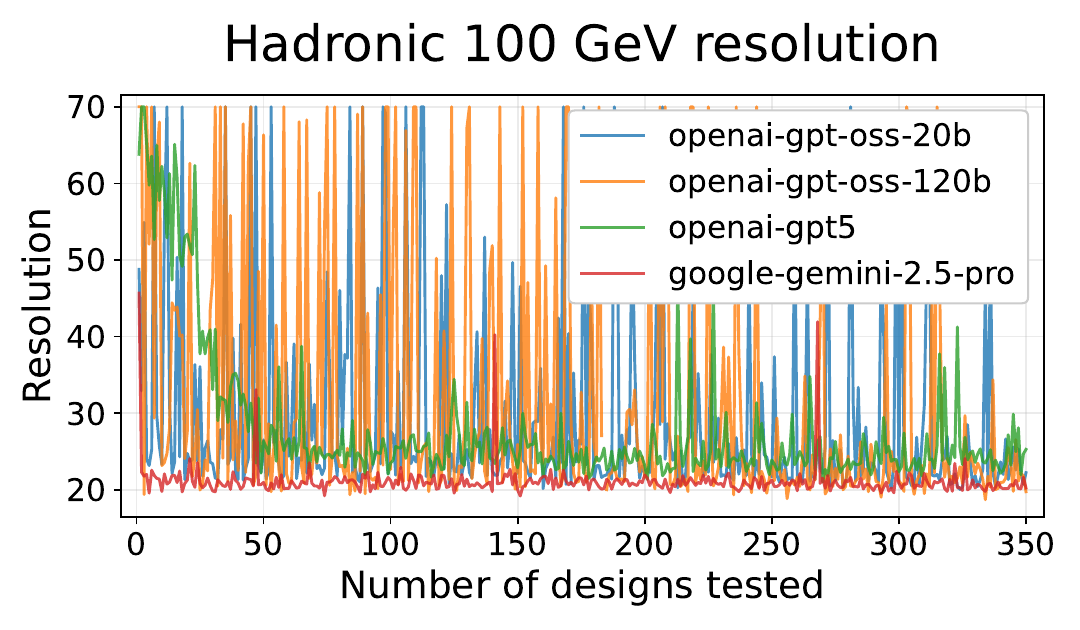}
        % \caption{Caption 4}
    \end{subfigure}
    \caption{Calorimeter performance across 350 design iterations. Energy resolution versus design iteration for multiple LLMs (GPT-OSS-20B, GPT-OSS-120B, GPT-5, and Gemini 2.5 Pro) over 350 iterations. The panels correspond to electromagnetic showers at 50 GeV (top left) and 100 GeV (top right), and hadronic showers at 50 GeV (bottom left) and 100 GeV (bottom right).}
    % [For LLMs for qualitative text generation: Results show quick "getting to" good results for GPT-OSS-20B and Gemini 2.5 Pro. They still improve later on (around 200th iteration but don't mention it exactly). GPT-5 starts bad but converges to a good value at 100th iteration. GPT-OSS-20B fluctuates a lot and very often gives very bad results but it's not really problem cuz if does often generates good performance. Compare in the table. The performance isn't that far off.]
    \label{fig:calo_plots_resolutions_only_llms_350it}
\end{figure}

Figure~\ref{fig:calo_plots_resolutions_only_llms_350it} summarizes the evolution of calorimeter performance over 350 proposal iterations. The four panels show the best-so-far mean-corrected energy resolution as a function of iteration for electromagnetic showers at 50 and 100~GeV (top row) and hadronic showers at 50 and 100~GeV (bottom row). Each curve corresponds to one of the evaluated language models: GPT-OSS-20B, GPT-OSS-120B, GPT-5, and Gemini~2.5~Pro. Across all models, the optimization exhibits rapid early progress, indicating that the proposal-and-evaluate loop quickly identifies physically meaningful segmentation patterns under the imposed feasibility and projection constraints.

Despite notable differences in model size and training provenance, all LLMs are able to substantially improve upon the baseline equidistant design. In particular, improvements are observed not only in electromagnetic resolution, where the baseline is already strong, but more prominently in the hadronic channels, which dominate the scalar reward. The resolution-level plots further reveal distinct optimization dynamics: while some models converge smoothly toward stable configurations, others exhibit stronger fluctuations yet still manage to repeatedly propose competitive designs.

\begin{figure}[ht]
    \centering
    %\textbf{}\\[0.8em]
    \includegraphics[page=1,width=0.95\textwidth]{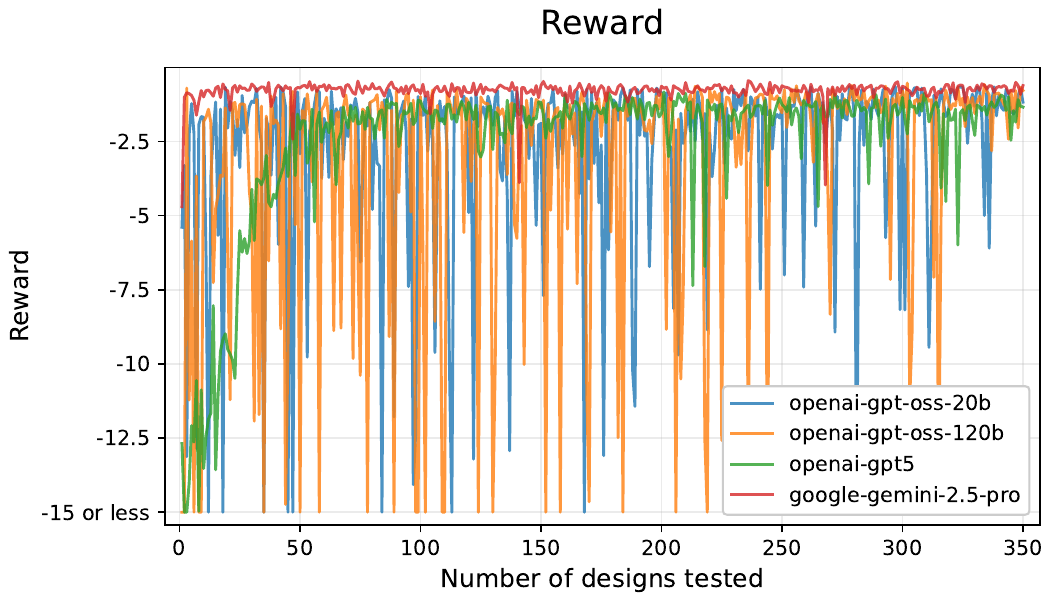}
    \caption{Calorimeter scalar reward as a function of design iteration for multiple reconstruction models (GPT-OSS-20B, GPT-OSS-120B, GPT-5, and Gemini 2.5 Pro), shown over 350 optimization iterations.}
    % [For LLMs for qualitative text generation: Similar as resolution performance plots. There is fluctuation in GPT-OSS-20B but has good results in general. GPT-5 converges to a good value around 100 iteration. Avoid giving exact iteration in the text generation.]
    \label{fig:calo_plot_reward_only_llms_350it}
\end{figure}

The corresponding evolution of the scalar reward is shown in Figure~\ref{fig:calo_plot_reward_only_llms_350it}. This figure aggregates the four resolution components into the single objective used by the evaluator and provides a compact summary of overall optimization progress. All models achieve rapid gains early in the optimization, followed by more gradual refinement. Differences in final reward values reflect trade-offs between electromagnetic and hadronic performance, consistent with the behavior observed in the resolution-level plots.

A quantitative comparison of the final designs is presented in Table~\ref{tab:calo_results_1}. Among the LLM-based approaches, Gemini~2.5~Pro achieves the strongest overall performance, producing electromagnetic resolutions close to those of the RL-optimized design and noticeably improved hadronic resolutions relative to the baseline. GPT-5 and GPT-OSS-120B also generate competitive configurations but do not surpass Gemini~2.5~Pro on this benchmark, particularly in the hadronic channels.

Most notably, GPT-OSS-20B performs remarkably well given its comparatively small size and fully open-weight nature. As shown in Table~\ref{tab:calo_results_1}, its hadronic energy resolutions at both 50 and 100~GeV represent a substantial improvement over the baseline design and approach the performance of significantly larger proprietary models. While its electromagnetic resolution remains slightly worse than that of the best-performing LLMs and the RL design, the overall performance gap is modest. The increased variability observed during optimization does not prevent GPT-OSS-20B from consistently discovering high-quality calorimeter configurations. 

\renewcommand{\arraystretch}{1.2} % only applies within this group
{
\begin{table}[ht]
\centering
\begin{tabular}{|l|l|l|l|l|}
\hline
\textbf{} & \textbf{50 GeV EM} & \textbf{100 GeV EM} & \textbf{50 GeV Had} & \textbf{100 GeV Had} \\ \hline
\textbf{Baseline design} & $8.15 \pm 0.16$  & $5.96 \pm 0.12$  & $32.13 \pm 0.64$  & $25.19 \pm 0.50$  \\\hline
\textbf{RL design} & $7.91 \pm 0.16$  & $5.65 \pm 0.11$  & $24.29 \pm 0.49$  & $18.07 \pm 0.36$  \\ \hline
\textbf{gpt-oss-20b} & $8.51 \pm 0.17$  & $6.15 \pm 0.12$  & $27.60 \pm 0.55$  & $20.22 \pm 0.40$  \\ \hline
\textbf{gpt-oss-20b + TR} & $8.38 \pm 0.17$  & $6.20 \pm 0.12$  & $26.39 \pm 0.53$  & $19.21 \pm 0.38$  \\ \hline
\textbf{gpt-oss-120b} & $9.39 \pm 0.19$  & $6.69 \pm 0.13$  & $27.69 \pm 0.55$  & $19.76 \pm 0.40$  \\ \hline
\textbf{gpt-oss-120b + TR} & $6.29 \pm 0.13$  & $4.54 \pm 0.09$  & $48.45 \pm 0.97$  & $51.58 \pm 1.03$  \\ \hline
\textbf{gpt-5} & $8.51 \pm 0.17$  & $6.28 \pm 0.13$  & $28.98 \pm 0.58$  & $21.11 \pm 0.42$  \\ \hline
\textbf{gpt-5 + TR} & $8.30 \pm 0.17$  & $6.14 \pm 0.12$  & $25.49 \pm 0.51$  & $19.18 \pm 0.38$  \\ \hline
\textbf{gemini-2.5-pro} & $7.98 \pm 0.16$  & $5.86 \pm 0.11$  & $26.57 \pm 0.53$  & $20.09 \pm 0.40$  \\ \hline
\textbf{gemini-2.5-pro + TR} & $8.04 \pm 0.16$  & $5.68 \pm 0.11$  & $25.09 \pm 0.50$  & $18.06 \pm 0.36$  \\ \hline

%\textbf{gemini-2.5flash} & $8.07 \pm 0.16$  & $6.09 \pm 0.12$  & $50.47 \pm 1.01$  & $46.6 \pm 0.93$  \\ \hline
%\textbf{claude-sonnet-4.5} & $8.81 \pm 0.18$  & $6.41 \pm 0.13$  & $33.34 \pm 0.67$  & $26.96 \pm 0.54$  \\ \hline
\end{tabular}
\caption{Calorimeter benchmark results. Energy resolution (in percent) for electromagnetic showers at 50 GeV and 100 GeV, and hadronic showers at 50 GeV and 100 GeV. Results are shown for the baseline design with equidistant sensors, the RL–optimized design, and designs generated by different LLMs (GPT-OSS-20B, GPT-OSS-120B, GPT-5, and Gemini~2.5~Pro). The LLM + TR row corresponds to the hybrid approach in which the LLM-proposed layout is refined by a $z$-only TR optimizer while keeping discrete design choices fixed. Quoted uncertainties correspond to statistical uncertainties from the evaluation samples.}
\label{tab:calo_results_1}
\end{table}
}

\begin{figure}[H] %[ht]
    \centering

    \begin{subfigure}[t]{0.49\textwidth}
        \centering
        \includegraphics[page=1,width=\textwidth]{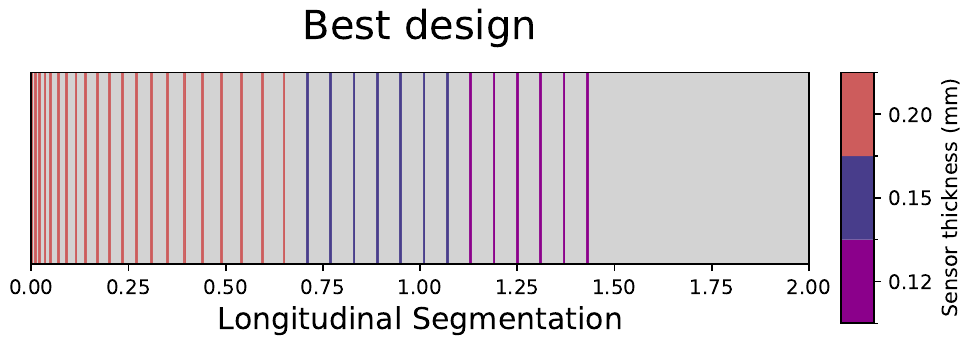}
        %\caption{GPT-OSS-20B}
        \label{fig:calo_best_20b}
    \end{subfigure}\hfill
    \begin{subfigure}[t]{0.49\textwidth}
        \centering
        \includegraphics[page=1,width=\textwidth]{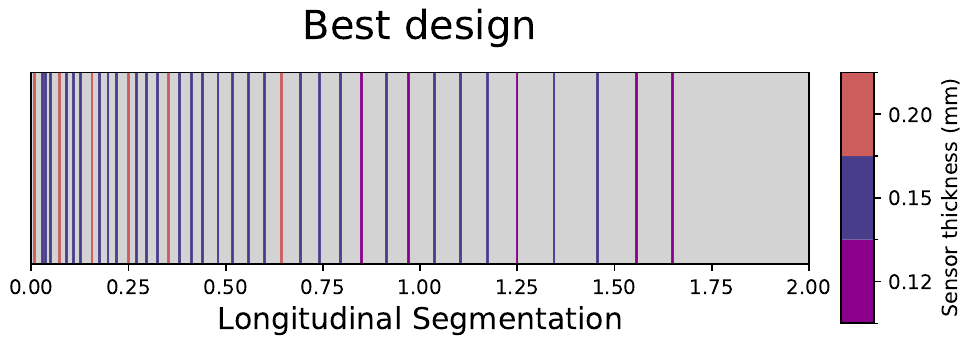}
        %\caption{Gemini 2.5 Pro}
        \label{fig:spectro_best_gemini_tr}
    \end{subfigure}

    \caption{Calorimeter best design found during different intervals during the training process for a standalone LLM (GPT-OSS-20B) on the left and a LLM (Gemini 2.5 Pro) + TR refinement on the right.}
    \label{fig:calo_best_all}
\end{figure}

\begin{figure} [H] %[ht]
    \centering    
    % Top-left
    \begin{subfigure}{0.48\textwidth}
        \centering
        \includegraphics[width=\textwidth]{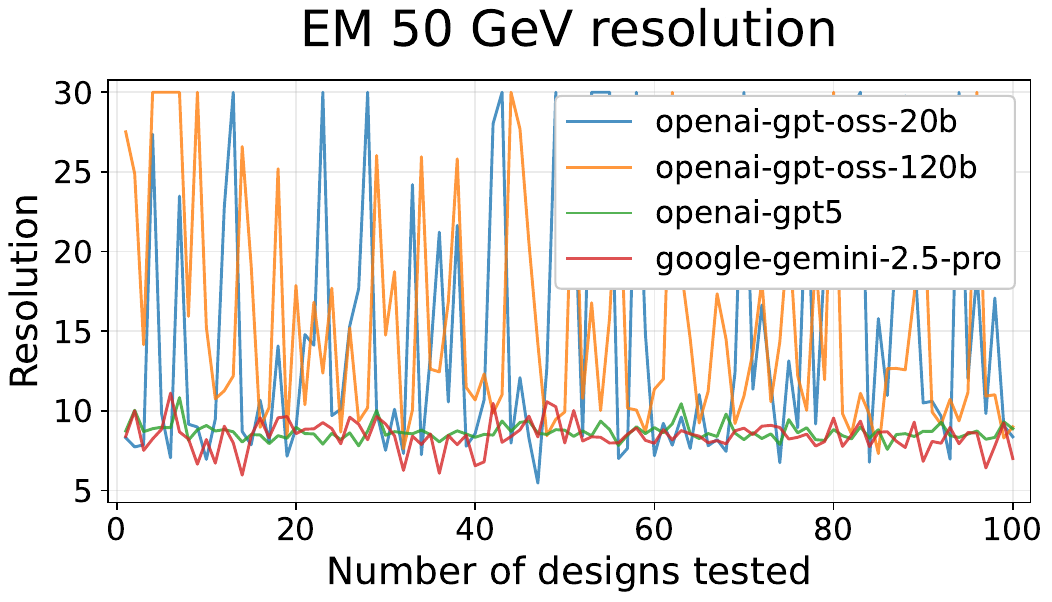} 
    \end{subfigure}
    \hfill
    % Top-right
    \begin{subfigure}{0.48\textwidth}
        \centering
        \includegraphics[width=\textwidth]{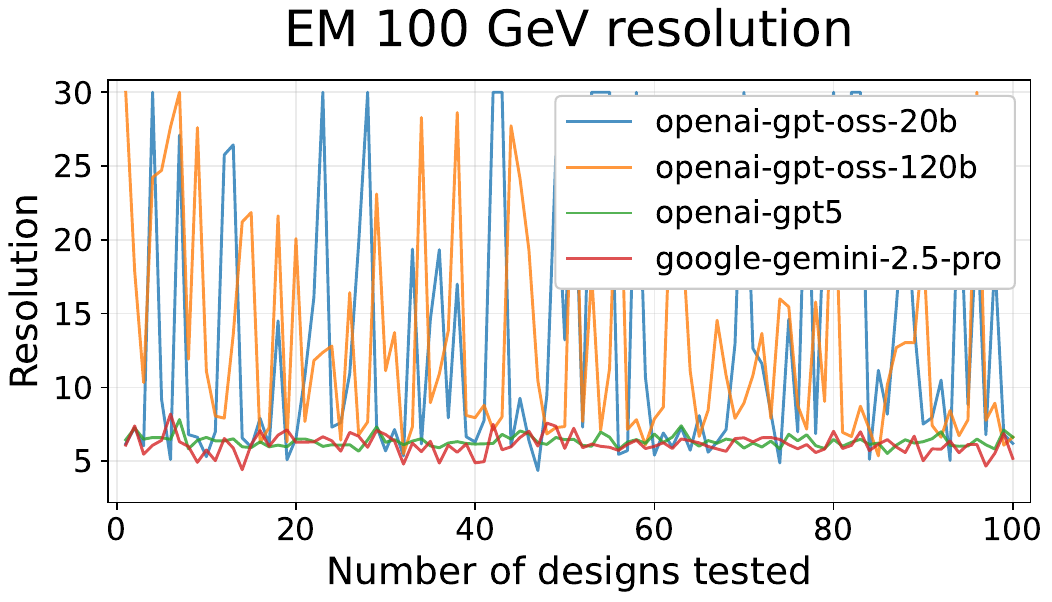}
    \end{subfigure}

    \vspace{0.5em}

    % Bottom-left
    \begin{subfigure}{0.48\textwidth}
        \centering
        \includegraphics[width=\textwidth]{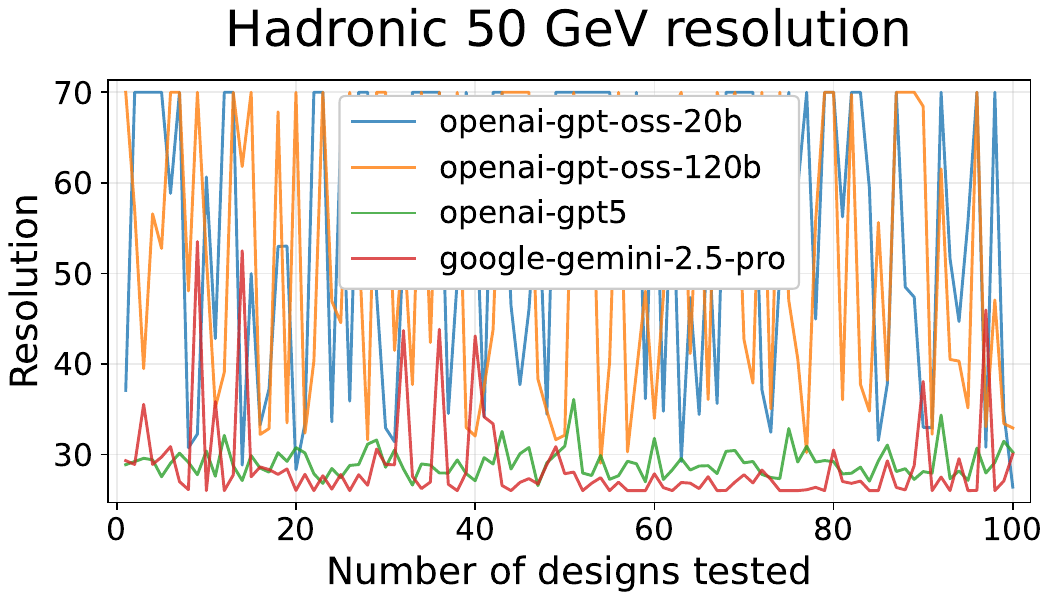}
    \end{subfigure}
    \hfill
    % Bottom-right
    \begin{subfigure}{0.48\textwidth}
        \centering
        \includegraphics[width=\textwidth]{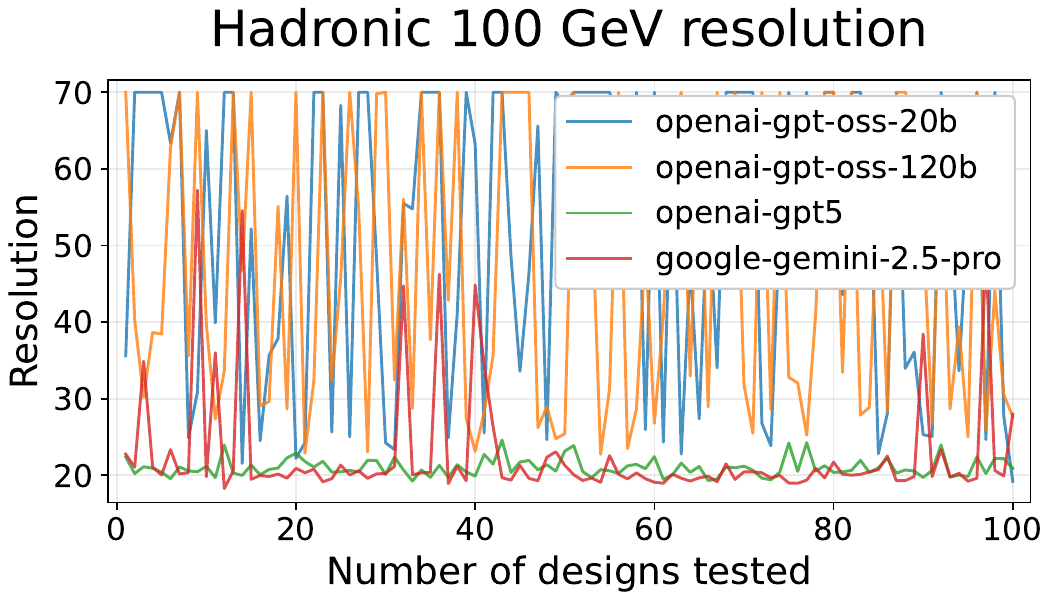}
    \end{subfigure}
    \caption{Calorimeter performance across design iterations. Tracking efficiency (left panels) and momentum resolution (right panels) as a function of design iteration over 100 proposal steps ($\times 100$ for TR refinement). The top row corresponds to 10 GeV tracks, and the bottom row to 100 GeV tracks. Curves are shown for the evaluated language models (GPT-OSS-20B, GPT-OSS-120B, GPT-5, and Gemini 2.5 Pro).}
    \label{fig:calo_performance_plots_bobyqa}
    % [For LLM qualitative text generation: All curves are similar to one without TR but show small improvements. 20B is the best one so far also confirmed by the numbers in the table.]
\end{figure}

\begin{figure}[H] %[ht]
    \centering
    %\textbf{}\\[0.8em]
    \includegraphics[page=1,width=0.95\textwidth]{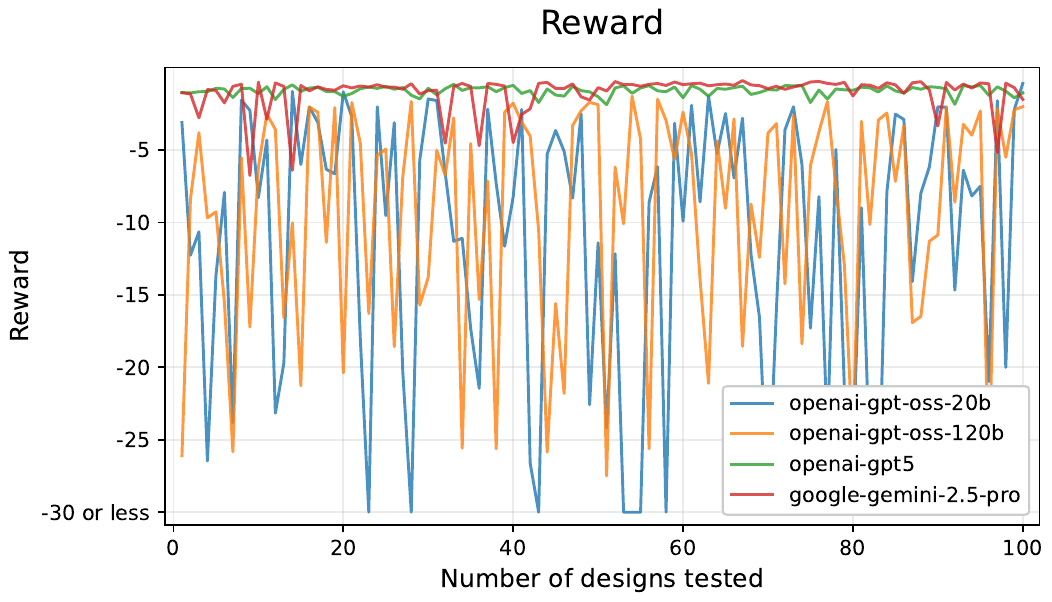}
    \caption{Calorimeter scalar reward as a function of design iteration over 100 proposal steps ($\times 100$ for TR refinement). The reward aggregates the spectrometer performance metrics into a single objective. Curves are shown for the evaluated language models (GPT-OSS-20B, GPT-OSS-120B, GPT-5, and Gemini 2.5 Pro).}
    \label{fig:calo_plot_reward_bobyqa}
    % [For LLM qualitative text generation: All curves are similar to one without TR but show small improvements. 20B is the best one so far also confirmed by the numbers in the table.]
\end{figure}

In addition to the standalone pipeline, we evaluate the hybrid variant introduced in Section~\ref{sec:methodology}, in which each LLM proposal is followed by a TR optimizer, with refinement restricted to the longitudinal placement variables $z$ while holding the other parameters fixed. We next quantify the impact of the hybrid pipeline in which each LLM proposal is followed by TR refinement on the $z$ coordinates. Figure~\ref{fig:tr_example} illustrates representative refinement trajectories: starting from the projected LLM design, the TR stage typically yields a consistent improvement over the initial reward and can recover additional performance without altering the discrete layer thickness choices.
Figure~\ref{fig:calo_performance_plots_bobyqa} shows the resulting best-so-far energy resolutions for electromagnetic and hadronic showers, and Figure~\ref{fig:calo_plot_reward_bobyqa} summarizes the corresponding scalar reward. Starting from a projected LLM proposal, the TR improves with respect to the LLM. Consistent with the summary in Table~\ref{tab:calo_results_1}, TR refinement provides modest but measurable gains for several models by fine-tuning layer boundaries without changing the discrete layer composition. At the same time, the behavior depends on the specific model and its initialization: for example, the GPT-OSS-120B + TR best design improves electromagnetic resolution but strongly degrades hadronic resolution, illustrating that $z$-only local refinement cannot compensate for suboptimal discrete choices and can amplify trade-offs between channels when the starting configuration is poorly matched to the full objective.

Figure~\ref{fig:calo_best_all} provides a qualitative illustration of the best calorimeter layout found at different stages of the run for a standalone LLM and for an LLM+TR pipeline. In both cases, the best designs evolve toward non-uniform, structured longitudinal segmentations rather than evenly spaced layers.

% Some stuff from here can later go in the outlook section: This result is particularly important for the broader scope of this work. It demonstrates that effective LLM-based planning for detector design does not require the largest available models, and that lightweight, open-source language models can already encode useful inductive biases for physically grounded design tasks. This opens the door to more accessible, transparent, and resource-efficient workflows for scientific instrument optimization.

% \begin{figure}[ht]
%     \centering
%     %\textbf{}\\[0.8em]
%     \includegraphics[page=1,width=0.95\textwidth]{figures/LLM/best_design/best_design_calo_20b_1.pdf}
%     \caption{Calorimeter best design found during different intervals during the training process of GPT-OSS-20B.}
%     \label{fig:calo_best}
% \end{figure}

% \begin{figure}[ht]
%     \centering
%     %\textbf{}\\[0.8em]
%     \includegraphics[page=1,width=0.95\textwidth]{figures/LLM/best_design/best_design_calo_gemini_tr.pdf}
%     \caption{Calorimeter best design found during different intervals during the training process of Gemini 2.5 Pro.}
%     \label{fig:calo_best_tr_2}
% \end{figure}

\subsection{Spectrometer}
\label{sec:results_spectro}
In the spectrometer benchmark, we evaluate the same LLM-based proposal-and-evaluate procedure on the tracking-station placement and granularity task introduced in Section~\ref{sec:method_spectro}. As in the calorimeter case, the simulator, reconstruction, and score definition are unchanged relative to the RL baseline; the only difference is the mechanism used to generate candidate designs. In addition to the standalone proposal loop, we also evaluate the hybrid variant in which each accepted LLM proposal is refined with the downstream TR stage described in Section~\ref{sec:methodology}, with refinement restricted to the longitudinal station positions while keeping the discrete granularity assignments fixed.

\begin{figure}[ht]
    \centering    
    % Top-left
    \begin{subfigure}{0.48\textwidth}
        \centering
        \includegraphics[width=\textwidth]{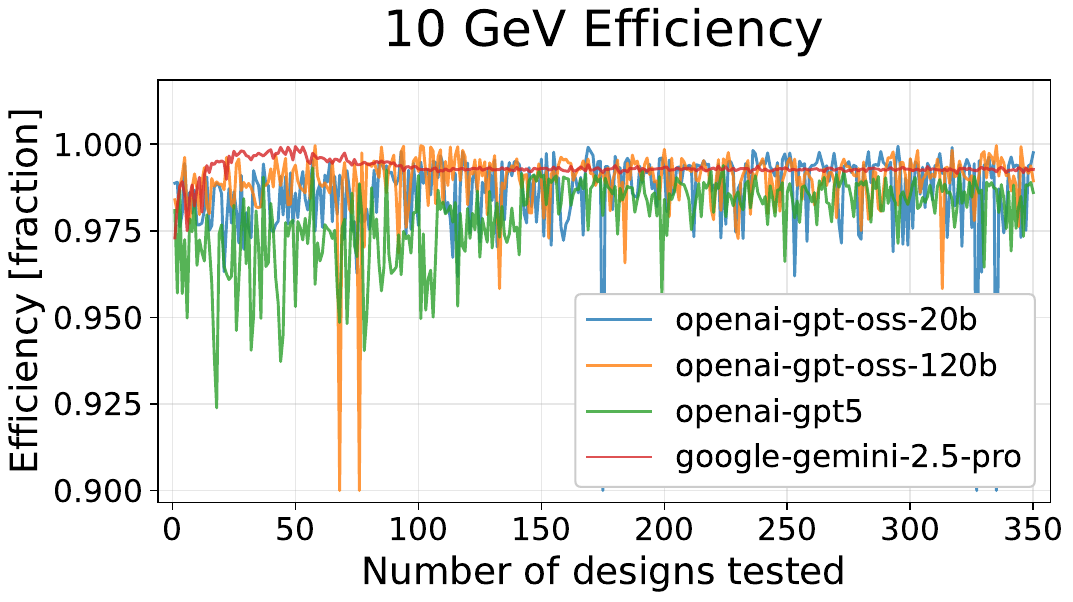} 
    \end{subfigure}
    \hfill
    % Top-right
    \begin{subfigure}{0.48\textwidth}
        \centering
        \includegraphics[width=\textwidth]{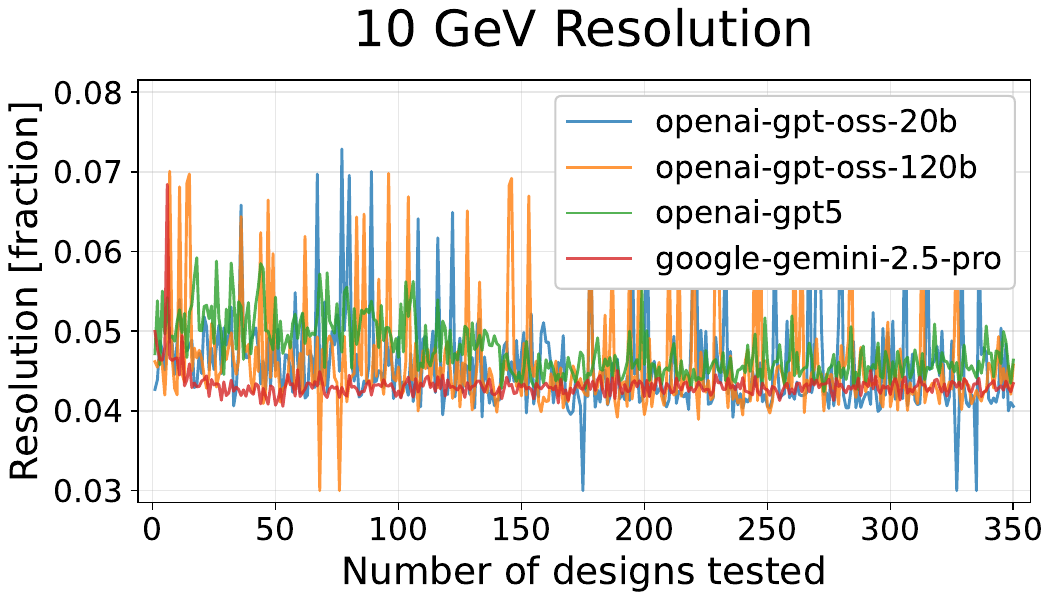}
    \end{subfigure}

    \vspace{0.5em}

    % Bottom-left
    \begin{subfigure}{0.48\textwidth}
        \centering
        \includegraphics[width=\textwidth]{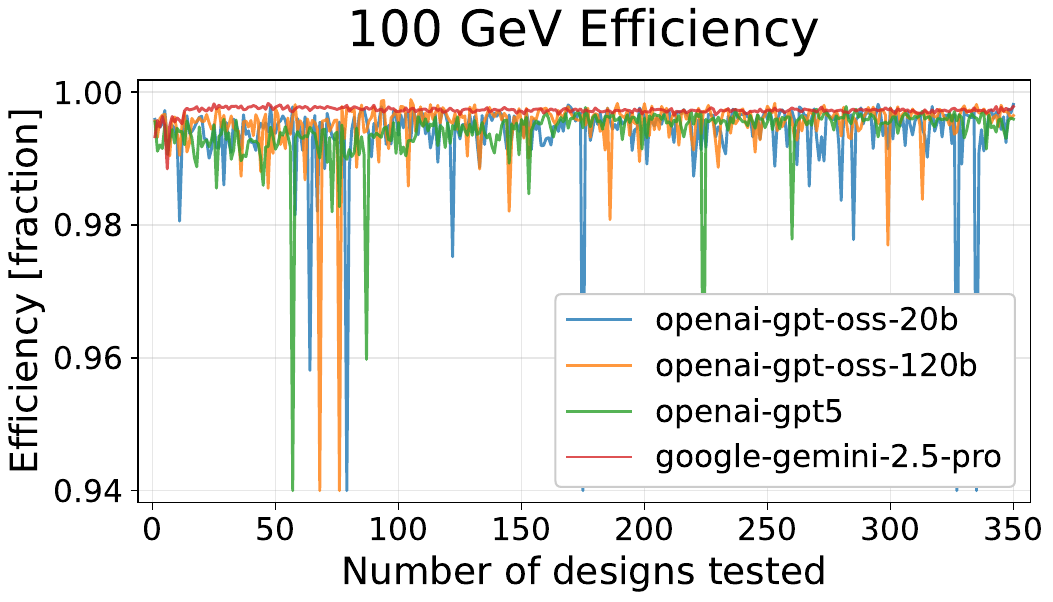}
    \end{subfigure}
    \hfill
    % Bottom-right
    \begin{subfigure}{0.48\textwidth}
        \centering
        \includegraphics[width=\textwidth]{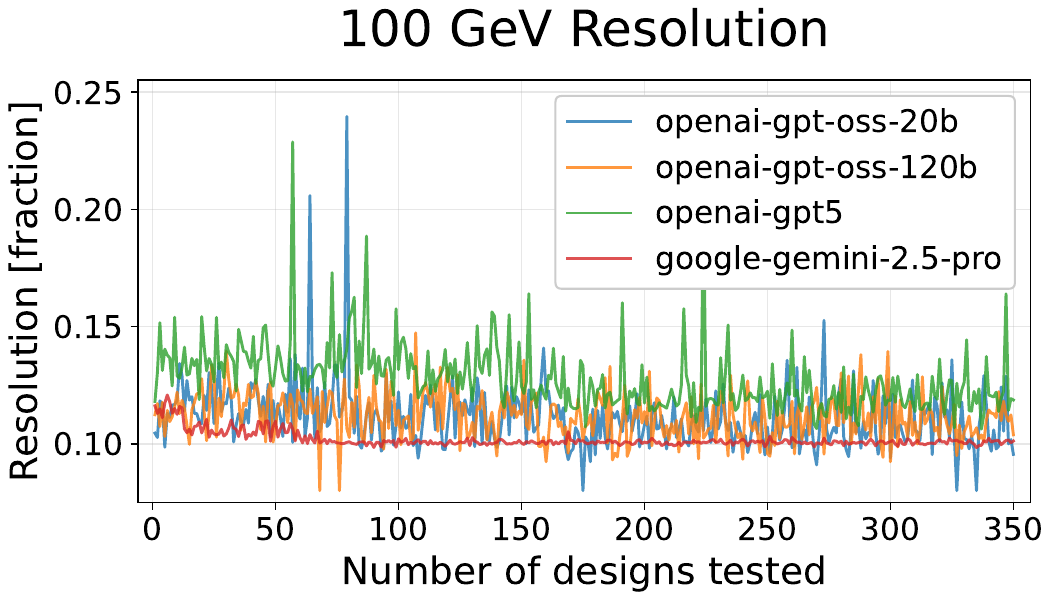}
    \end{subfigure}
    \caption{Spectrometer performance across 350 design iterations. Tracking efficiency (left panels) and momentum resolution (right panels) as a function of design iteration over 350 proposal steps. The top row corresponds to 10 GeV tracks, and the bottom row to 100 GeV tracks. Curves are shown for the evaluated language models (GPT-OSS-20B, GPT-OSS-120B, GPT-5, and Gemini 2.5 Pro).}
    \label{fig:spectro_performance_plots_only_llms_350it}
    % [For LLM qualitative text generation: Look at the figure below]
\end{figure}

Figure~\ref{fig:spectro_performance_plots_only_llms_350it} shows the best-so-far evolution of the two underlying physics metrics that enter the spectrometer score: tracking efficiency (left panels) and momentum resolution (right panels), evaluated at 10~GeV (top row) and 100~GeV (bottom row). Across all tested models, the proposal loop rapidly discovers physically reconstructable layouts after feasibility projection, and the best-so-far curves improve early before transitioning into slower, incremental refinement. The most visible gains occur at 10~GeV, where multiple scattering and limited lever arm make layout choices particularly consequential; nonetheless, improvements are also apparent at 100~GeV, indicating that the LLMs do not merely optimize a single regime but instead identify broadly effective station patterns.

\begin{figure}[ht]
    \centering
    %\textbf{}\\[0.8em]
    \includegraphics[page=1,width=0.95\textwidth]{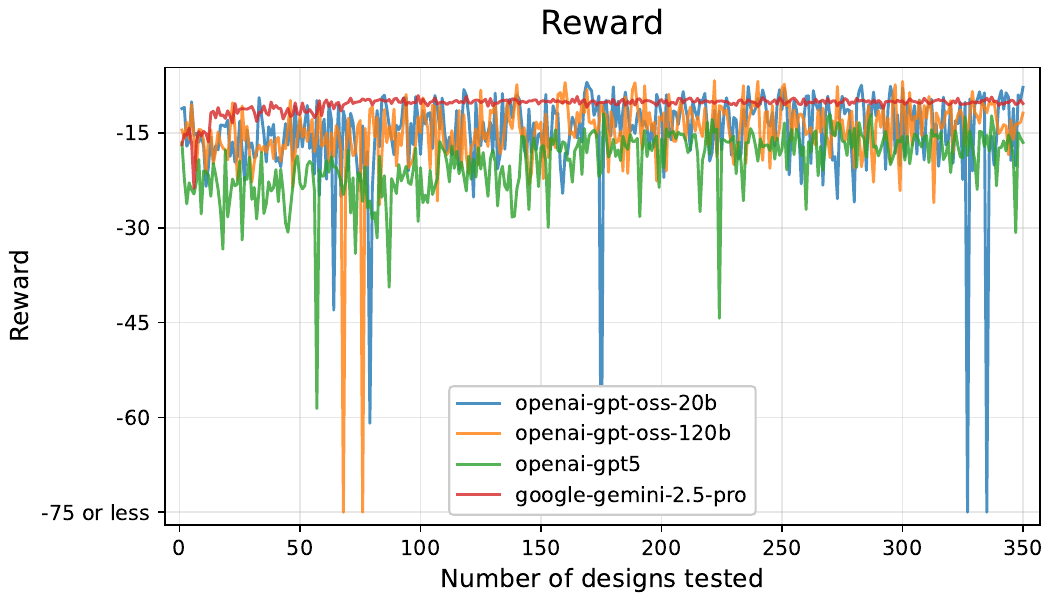}
    \caption{Spectrometer scalar reward as a function of design iteration over 350 proposal steps. The reward aggregates the spectrometer performance metrics into a single objective. Curves are shown for the evaluated language models (GPT-OSS-20B, GPT-OSS-120B, GPT-5, and Gemini 2.5 Pro).}
    \label{fig:spectro_plot_reward_only_llms_350it}
    % [For LLM qualitative text generation: GPT 5 fluctuates a lot but does deliver good performance it converges to a decent value around 50th iteration. OSS-20B fluctuates less but still a bit. Overall it performs the best. Gemini has close to no fluctuations. It converges to a good value around 70th iteration and then remains very constant and eventually 20B beats it.]
\end{figure}

The corresponding scalar objective is summarized in Figure~\ref{fig:spectro_plot_reward_only_llms_350it}. The curves highlight distinct search dynamics across models. Gemini~2.5~Pro exhibits stable optimization behavior, with relatively smooth progress and minimal fluctuations once a competitive configuration is found. In contrast, GPT-5 shows stronger variability in the proposed designs: while it occasionally produces weaker candidates, it also repeatedly recovers strong solutions and reaches a competitive best-so-far reward after the early exploration phase. The open-weight GPT-OSS-20B is particularly notable: despite its smaller scale, it consistently proposes high-quality configurations and ultimately matches or exceeds the best-so-far reward of the larger proprietary models within the same proposal budget. Overall, these trends reinforce the picture already seen in the calorimeter benchmark: LLMs act as effective generators of plausible detector layouts under mixed discrete--continuous constraints, with model-dependent trade-offs between stability and exploration.

\begin{figure}[ht]
    \centering

    \begin{subfigure}[t]{0.49\textwidth}
        \centering
        \includegraphics[page=1,width=\textwidth]{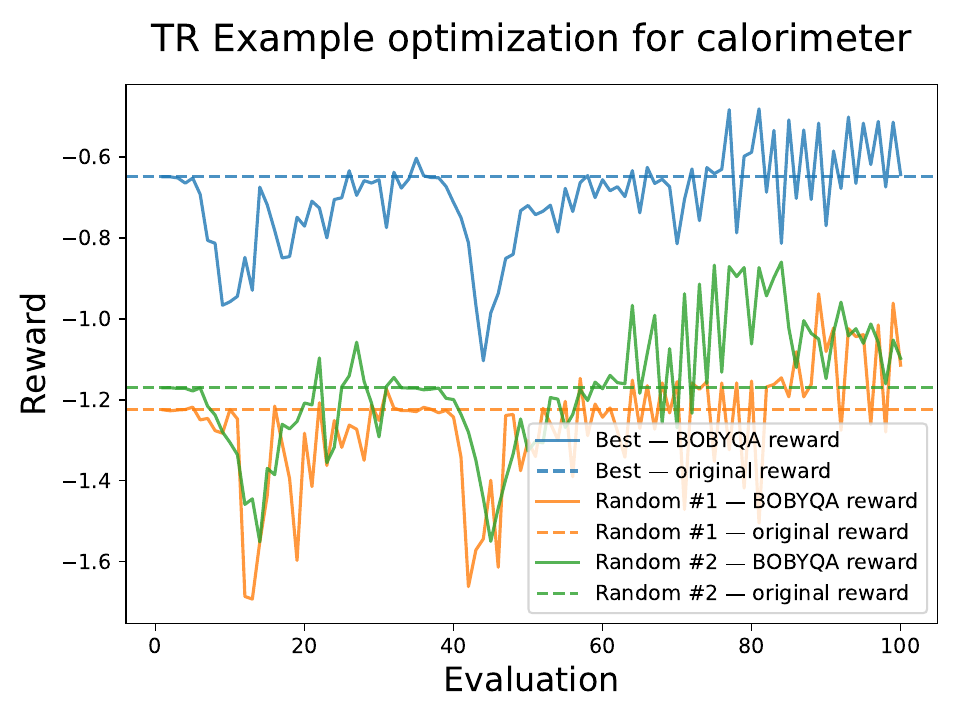}
        \label{fig:tr_example_calo}
    \end{subfigure}\hfill
    \begin{subfigure}[t]{0.49\textwidth}
        \centering
        \includegraphics[page=1,width=\textwidth]{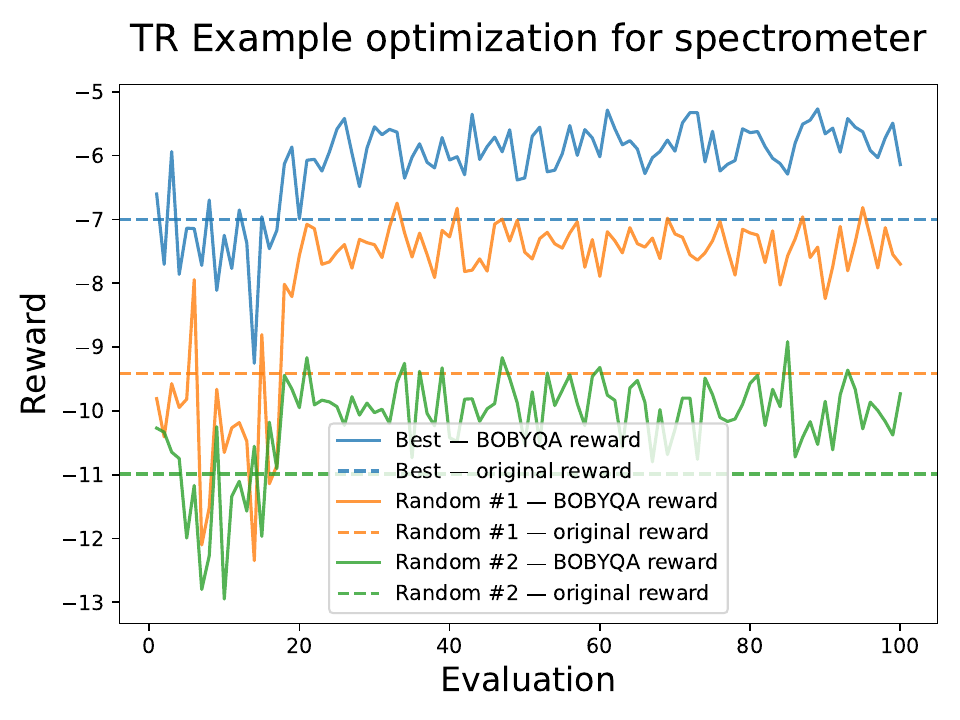}
        \label{fig:tr_example_spectro}
    \end{subfigure}

    \caption{TR Example Optimization Runs for the calorimeter (on the left) and the spectrometer (on the right). Three different colors show three designs (one is the best design and two are two random designs). It shows how the performance improves after applying the dedicated TR optimizer. There are also three horizontal lines giving the baseline reward from the initial design suggested by the LLM in same three colors as the plots for comparison.}
    \label{fig:tr_example}
\end{figure}

\begin{figure}[H] %[ht]
    \centering    
    % Top-left
    \begin{subfigure}{0.48\textwidth}
        \centering
        \includegraphics[width=\textwidth]{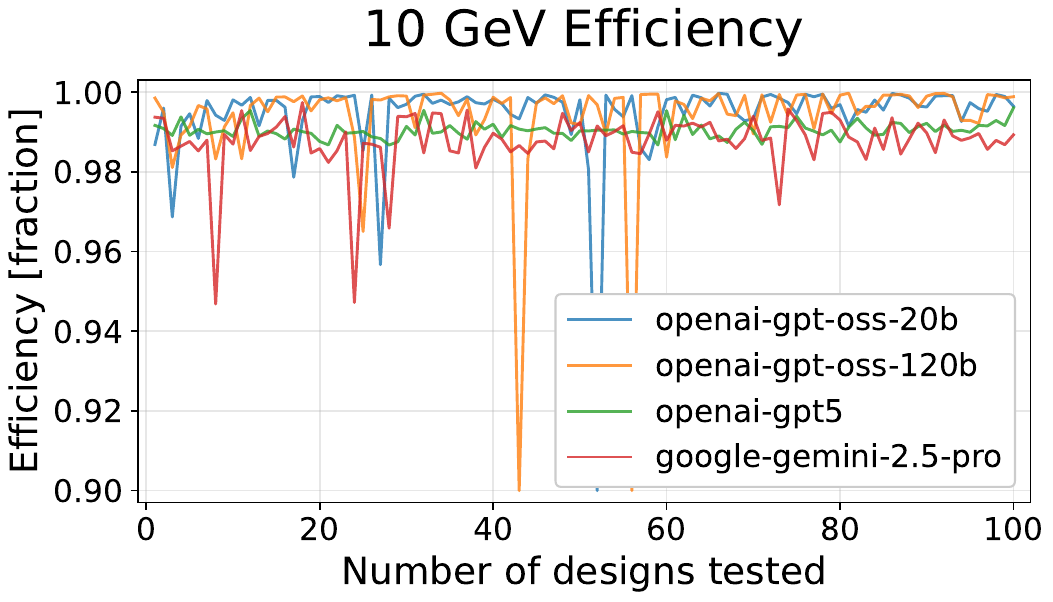} 
    \end{subfigure}
    \hfill
    % Top-right
    \begin{subfigure}{0.48\textwidth}
        \centering
        \includegraphics[width=\textwidth]{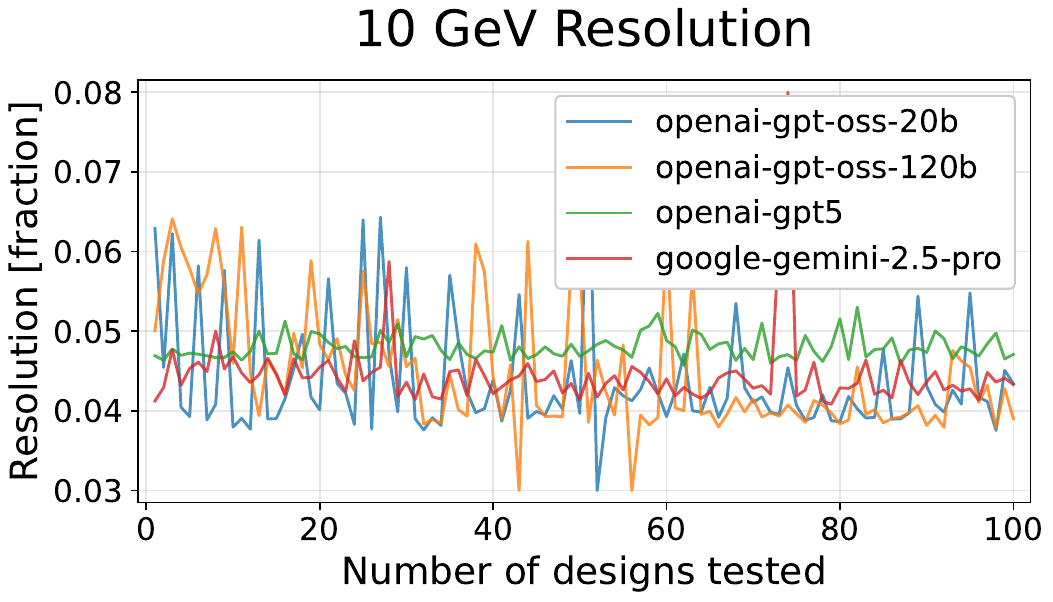}
    \end{subfigure}

    \vspace{0.5em}

    % Bottom-left
    \begin{subfigure}{0.48\textwidth}
        \centering
        \includegraphics[width=\textwidth]{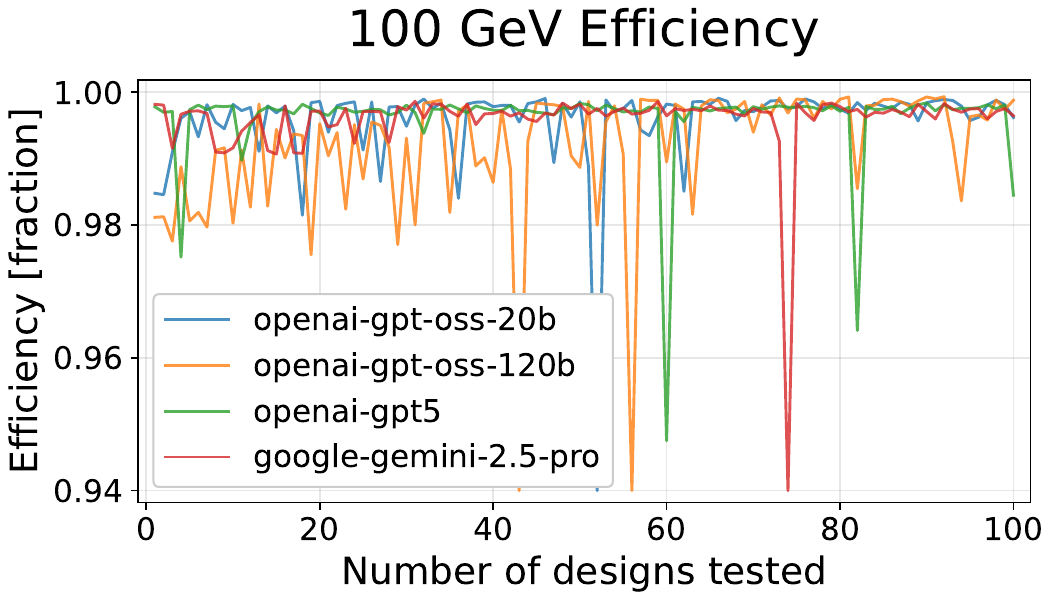}
    \end{subfigure}
    \hfill
    % Bottom-right
    \begin{subfigure}{0.48\textwidth}
        \centering
        \includegraphics[width=\textwidth]{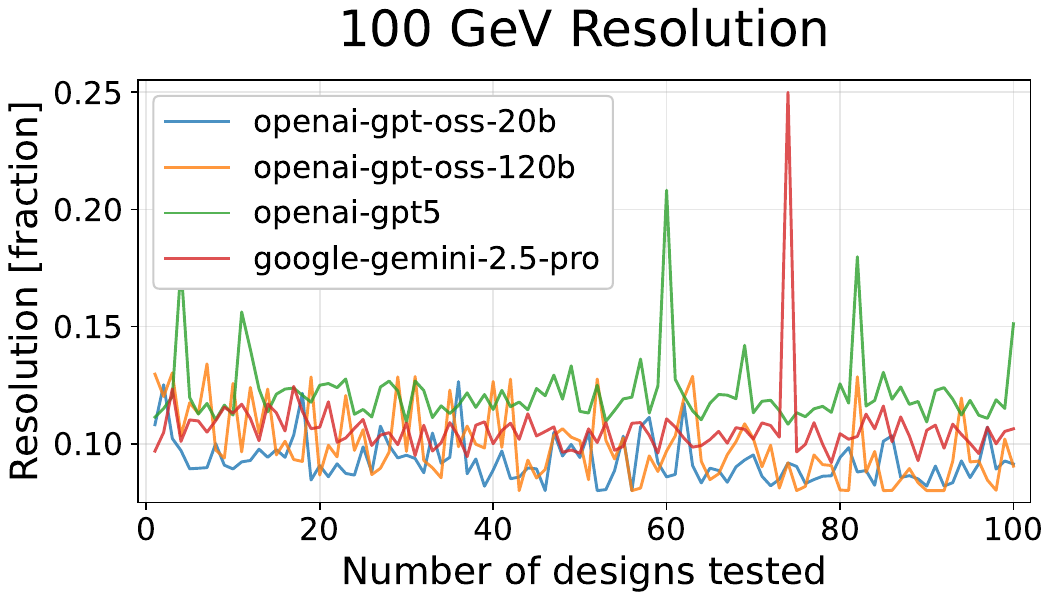}
    \end{subfigure}
    \caption{Spectrometer performance across design iterations. Tracking efficiency (left panels) and momentum resolution (right panels) as a function of design iteration over 100 proposal steps ($\times 100$ for TR refinement). The top row corresponds to 10 GeV tracks, and the bottom row to 100 GeV tracks. Curves are shown for the evaluated language models (GPT-OSS-20B, GPT-OSS-120B, GPT-5, and Gemini 2.5 Pro).}
    \label{fig:spectro_performance_plots_bobyqa}
    % [For LLM qualitative text generation: All curves are similar to one without TR but show small improvements. 20B is the best one so far also confirmed by the numbers in the table.]
\end{figure}

\begin{figure}[H] %[ht]
    \centering
    %\textbf{}\\[0.8em]
    \includegraphics[page=1,width=0.95\textwidth]{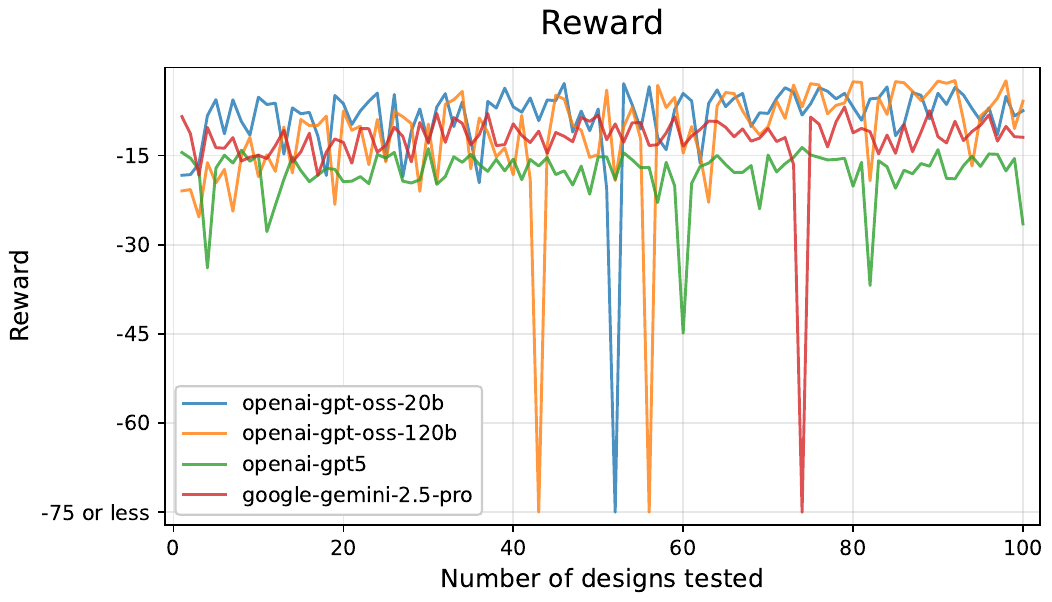}
    \caption{Spectrometer scalar reward as a function of design iteration over 100 proposal steps ($\times 100$ for TR refinement). The reward aggregates the spectrometer performance metrics into a single objective. Curves are shown for the evaluated language models (GPT-OSS-20B, GPT-OSS-120B, GPT-5, and Gemini 2.5 Pro).}
    \label{fig:spectro_plot_reward_bobyqa}
    % [For LLM qualitative text generation: All curves are similar to one without TR but show small improvements. 20B is the best one so far also confirmed by the numbers in the table.]
\end{figure}

We next quantify the impact of the hybrid pipeline in which each LLM proposal is followed by TR refinement on the $z$ coordinates. Figure~\ref{fig:tr_example} once again shows the representative refinement trajectories also for the spectrometer. Starting out from the projected LLM design, the TR stage yields a consistent improvement over the initial reward and recovers additional performance without altering the discrete station granularities. As for the calorimenter, Figure~\ref{fig:tr_example} shows a representative example. Aggregating over proposals, Figures~\ref{fig:spectro_performance_plots_bobyqa} and~\ref{fig:spectro_plot_reward_bobyqa} show that the resulting best-so-far curves shift upward relative to the standalone LLM runs, with the clearest gains appearing in momentum resolution. The effect is most pronounced when the initial proposal already places stations in a sensible pattern but leaves room for fine adjustment of lever arms and station spacing; in these cases, small longitudinal shifts translate into measurable improvements in resolution while maintaining near-saturated efficiencies.

A quantitative comparison of the best designs reached by each method is given in Table~\ref{tab:spectro_results}. All LLM-based approaches substantially improve over the baseline designs, achieving near-saturated efficiencies and significantly better momentum resolutions, especially at 10~GeV. Among the standalone LLMs, GPT-OSS-20B attains the strongest overall balance of efficiency and resolution, approaching the RL-optimized performance while remaining fully open-weight. Adding TR further reduces the gap to RL for the best-performing models: for GPT-OSS-20B and GPT-OSS-120B, TR refinement improves the momentum resolution and brings the 100~GeV resolution to essentially the RL level while preserving saturated efficiency. In contrast, the gains from TR are smaller and more model-dependent for GPT-5 and Gemini~2.5~Pro, reflecting that some proposed layouts are already near a local optimum in $z$ or that their remaining limitations are dominated by the fixed discrete granularity choices. 

\renewcommand{\arraystretch}{1.2} % only applies within this group
{
\begin{table}[H]%[ht]
\centering
\begin{tabular}{|l|l|l|l|l|}
\hline
\textbf{} & \textbf{10 GeV Eff} & \textbf{10 GeV Res} & \textbf{100 GeV Eff} & \textbf{100 GeV Res} \\ \hline
\textbf{Baseline design 1} & $88.86^{+0.25}_{-0.25}$ & $6.32 \pm 0.03$ & $98.03^{+0.11}_{-0.11}$ & $13.86 \pm 0.06$ \\ \hline
\textbf{Baseline design 2} & $93.60^{+0.19}_{-0.20}$ & $5.76 \pm 0.02$ & $99.17^{+0.07}_{-0.08}$ & $13.27 \pm 0.05$ \\ \hline
\textbf{RL design} & $100.00^{+0.00}_{-0.01}$ & $3.49 \pm 0.01$ & $99.90^{+0.02}_{-0.03}$ & $7.95 \pm 0.03$ \\ \hline
\textbf{gpt-oss-20b} & $99.91^{+0.01}_{-0.01}$ & $4.01 \pm 0.02$ & $99.81^{+0.02}_{-0.02}$ & $9.32 \pm 0.04$ \\ \hline
\textbf{gpt-oss-20b + TR} & $99.94^{+0.01}_{-0.01}$ & $3.95 \pm 0.02$ & $99.91^{+0.01}_{-0.01}$ & $7.97 \pm 0.03$ \\ \hline
\textbf{gpt-oss-120b} & $99.58^{+0.03}_{-0.03}$ & $4.05 \pm 0.02$ & $99.76^{+0.02}_{-0.02}$ & $9.23 \pm 0.04$ \\ \hline
\textbf{gpt-oss-120b + TR} & $99.95^{+0.01}_{-0.01}$ & $3.74 \pm 0.02$ & $99.87^{+0.02}_{-0.02}$ & $7.98 \pm 0.03$ \\ \hline
\textbf{gpt-5} & $99.02^{+0.04}_{-0.04}$ & $4.28 \pm 0.02$ & $99.77^{+0.02}_{-0.02}$ & $10.72 \pm 0.04$ \\ \hline
\textbf{gpt-5 + TR} & $99.11^{+0.04}_{-0.04}$ & $4.71 \pm 0.02$ & $99.78^{+0.02}_{-0.02}$ & $10.84 \pm 0.04$ \\ \hline
%\textbf{x-ai-grok-4-fast} & $95.29^{+0.09}_{-0.09}$ & $5.23 \pm 0.02$ & $98.37^{+0.05}_{-0.05}$ & $10.45 \pm 0.04$ \\ \hline
%\textbf{meta-llama-4-maverick} & $98.58^{+0.05}_{-0.05}$ & $6.33 \pm 0.03$ & $99.16^{+0.04}_{-0.04}$ & $9.70 \pm 0.04$ \\ \hline
\textbf{gemini-2.5-pro} & $99.35^{+0.03}_{-0.03}$ & $4.18 \pm 0.02$ & $99.71^{+0.02}_{-0.02}$ & $9.85 \pm 0.04$ \\ \hline
\textbf{gemini-2.5-pro + TR} & $99.50^{+0.03}_{-0.03}$ & $4.08 \pm 0.02$ & $99.84^{+0.02}_{-0.02}$ & $9.19 \pm 0.04$ \\ \hline
\end{tabular}
\caption{Spectrometer benchmark results. Tracking efficiency (in percent) and momentum resolution (in percent) evaluated at 10~GeV and 100~GeV for two baseline designs (respectively with 3 and 4 equidistant tracking stations, placed in both Region A and Region C, as described in \textcite{qasim2024physics-instrument-rl}), the RL–optimized (RL) design, and designs generated by LLMs (GPT-OSS-20B, GPT-OSS-120B, GPT-5, and Gemini~2.5~Pro). The LLM + TR row corresponds to the hybrid approach in which the LLM-proposed layout is refined by a $z$-only TR optimizer while keeping discrete design choices fixed. Quoted uncertainties are statistical uncertainties from the evaluation samples.}
\label{tab:spectro_results}
\end{table}
}

\begin{figure}[H] %[ht]
    \centering

    \begin{subfigure}[t]{0.49\textwidth}
        \centering
        \includegraphics[page=1,width=\textwidth]{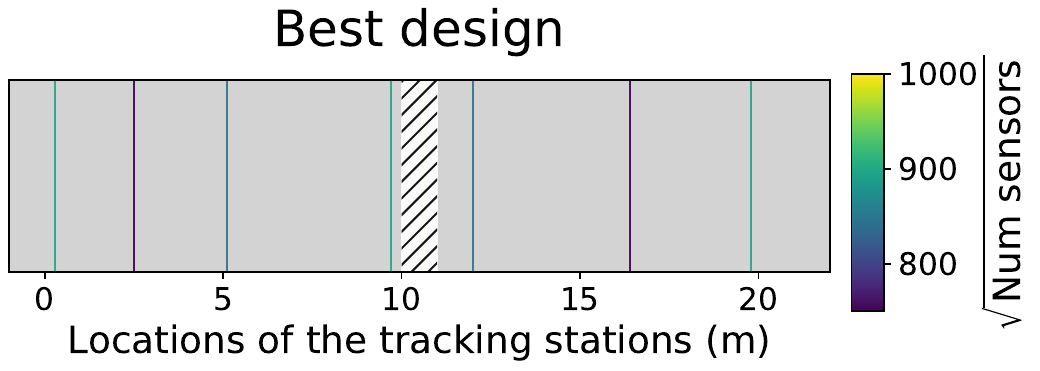}
        %\caption{GPT-5}
        \label{fig:spectro_best_gpt5}
    \end{subfigure}\hfill
    \begin{subfigure}[t]{0.49\textwidth}
        \centering
        \includegraphics[page=1,width=\textwidth]{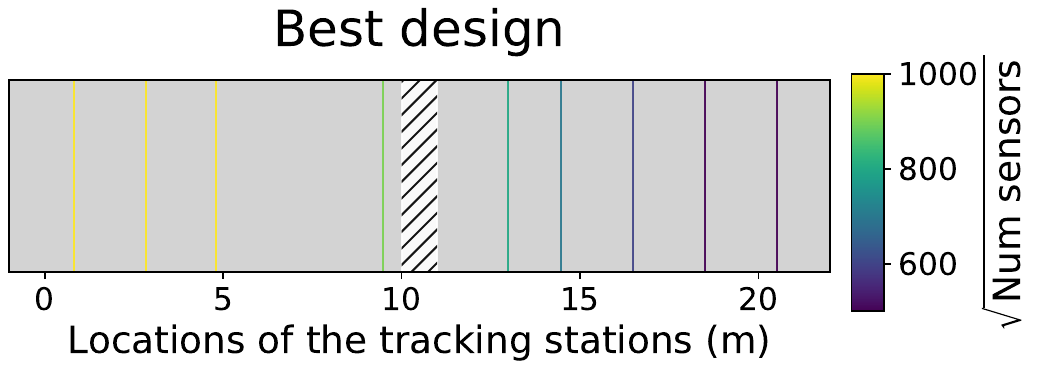}
        %\caption{GPT-OSS-20B}
        \label{fig:spectro_best_20b_tr}
    \end{subfigure}

    \caption{Spectrometer best design found during different intervals during the training process for a standalone LLM (GPT-5) on the left and a LLM (GPT-OSS-20B) + TR refinement on the right.}
    \label{fig:spectro_best_all}
\end{figure}

Figure~\ref{fig:spectro_best_all} provides a qualitative representation of the best station patterns found during the run evolve over time, both for a standalone LLM and for an LLM+TR run. The progression typically reflects increasingly sensible lever arms and spacing while staying within the global pixel budget. These results show that $z$-only TR refinement can extract additional performance from feasible LLM proposals, particularly by improving momentum resolution without changing the fixed granularity assignments.

\section{Discussion}
\label{sec:outlook}

The two benchmark studies show that LLMs can play a meaningful and effective role in physics instrument design even when used purely as proposal generators. Across both the calorimeter and spectrometer problems, LLMs—used without task-specific training and without direct interaction with the simulator—consistently generate physically meaningful, resource-aware detector configurations. Within a limited number of proposal iterations, these designs substantially outperform baseline layouts and recover a large fraction of the performance achieved by RL, while exhibiting sensible structural features. This behavior indicates that the LLM encodes useful inductive biases for detector design that can be exploited to guide structured exploration of the design space.

%The results of the two benchmark studies clarify a complementary role for large language models for design. In both the calorimeter and spectrometer design problems, RL ultimately achieves the best-performing designs. At the same time, LLMs---used without task-specific training and without direct simulator interaction---consistently generate physically meaningful and resource-aware detector configurations. Even without downstream refinement, these designs substantially outperform baseline layouts and exhibit sensible structural features, indicating that the LLM captures useful inductive biases for detector design.

We additionally introduced a trust-region (TR) refinement stage as a proof of concept for hybrid optimization pipelines. When applied downstream of LLM proposals, this lightweight continuous optimizer consistently improves performance and narrows the gap to the RL baseline. The remaining difference can be naturally explained by the limitations of the TR approach itself: it is restricted to continuous placement variables and cannot revise discrete design decisions, which are known to be critical in these benchmarks. In a realistic end-to-end pipeline, this refinement role would naturally be assumed by RL itself. 

The TR refinement experiments also help localize where performance improvements remain available. Because refinement is restricted to continuous placement variables ($z$) with discrete choices held fixed, the gains from TR can be interpreted as the “headroom” available from geometric fine-tuning alone. The fact that TR reliably improves performance yet does not fully close the gap to RL suggests that a meaningful fraction of the remaining improvement is driven by design decisions that are not accessible to $z$-only local search—most notably the discrete choices (layer types or station granularities) and, more generally, structural reconfigurations. This supports the hybrid view in which LLMs provide feasible, high-quality initial hypotheses while a reward-driven optimizer capable of revising discrete structure is needed to reach the best designs.

A notable practical observation is the speed at which LLMs begin to generate meaningful designs. In both benchmarks, physically sensible configurations appear within a few tens of proposal iterations, often well before the final best design is found. This suggests a realistic workflow in which LLM-based proposal generation is used to rapidly identify promising regions of the design space, after which RL can be deployed on a much smaller and better-structured search domain. At this scale, RL optimization runs become tractable even for complex detector layouts.

Similar hybrid roles for LLMs are emerging in other design and optimization settings, where language models provide problem- or domain-aware guidance while a dedicated optimizer remains responsible for objective-driven improvement. Examples include LLM-augmented Bayesian optimization, in which language models influence surrogate modeling or acquisition decisions \parencite{liu_llms_for_bayesian_2024,austin_llms_for_bayesian_2024}, and circuit design workflows in which an LLM orchestrates high-level topology selection while a separate numerical optimizer tunes continuous parameters \parencite{jin2025_wiseEDA}.

This supports a clear division of labor where RL remains the core optimization methodology for physics instrument design, capable of extracting maximum performance when reward-driven search is feasible. LLMs complement RL by providing structured, physically informed priors that reduce reward sparsity, limit wasted exploration, and automate parts of the human problem-structuring process that currently constrain scalability. This hybrid perspective offers a concrete path toward closed-loop, end-to-end detector design at the level of complexity required for next-generation experiments.

\section{Conclusion} \label{sec:conclusion}
In this work, we investigated the use of LLMs for physics instrument design using two representative benchmark problems: the longitudinal segmentation of a sampling calorimeter and the layout and granularity optimization of a magnetic spectrometer. The detector models, simulations, reconstruction algorithms, and reward definitions were kept identical to those used in established RL studies, ensuring that differences in performance arise solely from the design-generation strategy. The LLMs were used without task-specific training and interacted with the design loop only through structured design descriptions evaluated by the same physics-based metrics as the RL baselines.

Across both benchmarks, the LLMs consistently generated valid, resource-aware, and physically meaningful detector designs. Within a limited number of iterations, they substantially outperformed baseline configurations and recovered a large fraction of the performance achieved by RL. While RL remains the most effective approach for reaching the strongest final designs, the LLM-based designs demonstrate that pretrained models can encode useful inductive biases about detector layout, particle--matter interactions, and resource trade-offs, even when applied in a purely prompted setting.

We also explored a minimal hybrid variant in which LLM-generated designs were refined by a dedicated trust-region optimizer acting on continuous placement variables. This refinement stage reliably improved performance and further narrowed the gap to RL on the spectrometer benchmark, showing that LLM-generated designs can serve as strong starting points for downstream optimization without altering discrete structural choices.

Based on these studies, we argued for a role of LLMs not as replacements for physics-driven optimization, but as high-level contributors to the design process, providing structured, informed design hypotheses that reduce unproductive exploration and simplify the organization of optimization workflows. This perspective naturally points toward hybrid pipelines in which LLMs assist with structuring and coordinating design studies, while established optimization methods remain responsible for extracting maximal performance under well-defined physics objectives.

A natural next step is to replace the lightweight trust-region refinement used here with a full RL optimizer downstream of the LLM proposals. Unlike $z$-only TR tuning, RL can revise both continuous placements and discrete structural choices, and therefore has the capacity to recover the remaining headroom that is inaccessible to local continuous refinement. More broadly, an important direction is to move beyond optimizing a single subsystem in isolation and toward coordinated co-design of multiple detector components. In such a setting, the LLM can provide an actionable prior and manage separate optimization loops (e.g., dedicated RL agents or optimizers per subsystem), while reasoning about their coupling through a shared physics objective. For example, calorimetric and tracking measurements can provide complementary constraints on particle kinematics, so the globally optimal instrument may not correspond to independently optimal subsystem designs. This motivates hybrid pipelines in which the LLM explicitly proposes how to factorize the joint design problem, what information to exchange between subsystem optimizers, and how to allocate evaluation budget so that independent RL searches can be coordinated to achieve the desired end-to-end performance.

%\break
%\printbibliography
% \bibliographystyle{plainnat}
% \bibliography{references}
%\clearpage
%\FloatBarrier
\printbibliography

@misc{qasim2024physics-instrument-rl,
  title         = {Physics Instrument Design with Reinforcement Learning},
  author        = {Qasim, Shah Rukh and Owen, Patrick and Serra, Nicola},
  year          = {2024},
  eprint        = {2412.10237},
  archivePrefix = {arXiv},
  primaryClass  = {physics.ins-det},
  doi           = {10.48550/arXiv.2412.10237},
  url           = {https://arxiv.org/abs/2412.10237}
}

@article{mirhoseini2021graph,
  title = {Author Correction: A graph placement methodology for fast chip design},
  volume = {604},
  ISSN = {1476-4687},
  url = {http://dx.doi.org/10.1038/s41586-022-04657-6},
  DOI = {10.1038/s41586-022-04657-6},
  number = {7906},
  journal = {Nature},
  publisher = {Springer Science and Business Media LLC},
  author = {Mirhoseini,  Azalia and Goldie,  Anna and Yazgan,  Mustafa and Jiang,  Joe Wenjie and Songhori,  Ebrahim and Wang,  Shen and Lee,  Young-Joon and Johnson,  Eric and Pathak,  Omkar and Nova,  Azade and Pak,  Jiwoo and Tong,  Andy and Srinivasa,  Kavya and Hang,  William and Tuncer,  Emre and Le,  Quoc V. and Laudon,  James and Ho,  Richard and Carpenter,  Roger and Dean,  Jeff},
  year = {2022},
  month = mar,
  pages = {E24–E24}
}

@article{baranov2017optimising,
  title = {Optimising the Active Muon Shield for the {SHiP} Experiment at {CERN}},
  volume = {934},
  ISSN = {1742-6596},
  url = {http://dx.doi.org/10.1088/1742-6596/934/1/012050},
  DOI = {10.1088/1742-6596/934/1/012050},
  journal = {Journal of Physics: Conference Series},
  publisher = {IOP Publishing},
  author = {Baranov,  A and Burnaev,  E and Derkach,  D and Filatov,  A and Klyuchnikov,  N and Lantwin,  O and Ratnikov,  F and Ustyuzhanin,  A and Zaitsev,  A},
  year = {2017},
  month = dec,
  pages = {012050}
}

@inproceedings{shirobokov2020black,
 author = {Shirobokov, Sergey and Belavin, Vladislav and Kagan, Michael and Ustyuzhanin, Andrei and Baydin, Atilim Gunes},
 booktitle = {Advances in Neural Information Processing Systems},
 editor = {H. Larochelle and M. Ranzato and R. Hadsell and M.F. Balcan and H. Lin},
 pages = {14650--14662},
 publisher = {Curran Associates, Inc.},
 title = {Black-Box Optimization with Local Generative Surrogates},
 url = {https://proceedings.neurips.cc/paper_files/paper/2020/file/a878dbebc902328b41dbf02aa87abb58-Paper.pdf},
 volume = {33},
 year = {2020}
}

@article{geant4_cite,
  title = {Geant4—a simulation toolkit},
  volume = {506},
  ISSN = {0168-9002},
  url = {http://dx.doi.org/10.1016/S0168-9002(03)01368-8},
  DOI = {10.1016/s0168-9002(03)01368-8},
  number = {3},
  journal = {Nuclear Instruments and Methods in Physics Research Section A: Accelerators,  Spectrometers,  Detectors and Associated Equipment},
  publisher = {Elsevier BV},
  author = "{{GEANT4 Collaboration}}",
  year = {2003},
  month = jul,
  pages = {250–303}
}

@article{dorigo2023toward,
  title = {Toward the end-to-end optimization of particle physics instruments with differentiable programming},
  volume = {10},
  ISSN = {2405-4283},
  url = {http://dx.doi.org/10.1016/j.revip.2023.100085},
  DOI = {10.1016/j.revip.2023.100085},
  journal = {Reviews in Physics},
  publisher = {Elsevier BV},
  author = {Dorigo,  Tommaso and Giammanco,  Andrea and Vischia,  Pietro and Aehle,  Max and Bawaj,  Mateusz and Boldyrev,  Alexey and de Castro Manzano,  Pablo and Derkach,  Denis and Donini,  Julien and Edelen,  Auralee and Fanzago,  Federica and Gauger,  Nicolas R. and Glaser,  Christian and Baydin,  Atılım G. and Heinrich,  Lukas and Keidel,  Ralf and Kieseler,  Jan and Krause,  Claudius and Lagrange,  Maxime and Lamparth,  Max and Layer,  Lukas and Maier,  Gernot and Nardi,  Federico and Pettersen,  Helge E.S. and Ramos,  Alberto and Ratnikov,  Fedor and R\"{o}hrich,  Dieter and de Austri,  Roberto Ruiz and del Árbol,  Pablo Martínez Ruiz and Savchenko,  Oleg and Simpson,  Nathan and Strong,  Giles C. and Taliercio,  Angela and Tosi,  Mia and Ustyuzhanin,  Andrey and Zaraket,  Haitham},
  year = {2023},
  month = jun,
  pages = {100085}
}

@misc{mode_collaboration,
  author       = {{MODE Collaboration}},
  title        = {{Machine-learning Optimized Design of Experiments} ({MODE})},
  year         = 2024,
  url          = {https://mode-collaboration.github.io},
  note         = {Accessed: 2024-10-31}
}

@article{reoptimized_lhcb,
  title = {Overview of the {LHCb} experiment},
  volume = {446},
  ISSN = {0168-9002},
  url = {http://dx.doi.org/10.1016/S0168-9002(00)00014-0},
  DOI = {10.1016/s0168-9002(00)00014-0},
  number = {1–2},
  journal = {Nuclear Instruments and Methods in Physics Research Section A: Accelerators,  Spectrometers,  Detectors and Associated Equipment},
  publisher = {Elsevier BV},
  author = {Schneider,  Olivier},
  year = {2000},
  month = may,
  pages = {213–221}
}

@techreport{na62_status,
      author        = "{{NA62 Collaboration}}",
      collaboration = "NA62",
      title         = "{2022 NA62 Status Report to the CERN SPSC}",
      institution   = "CERN",
      reportNumber  = "CERN-SPSC-2022-012, SPSC-SR-306",
      address       = "Geneva",
      year          = "2022",
      url           = "https://cds.cern.ch/record/2805351",
}

@article{Strong2024,
  title = {{TomOpt}: differential optimisation for task- and constraint-aware design of particle detectors in the context of muon tomography},
  volume = {5},
  ISSN = {2632-2153},
  url = {http://dx.doi.org/10.1088/2632-2153/ad52e7},
  DOI = {10.1088/2632-2153/ad52e7},
  number = {3},
  journal = {Machine Learning: Science and Technology},
  publisher = {IOP Publishing},
  author = {Strong,  Giles C and Lagrange,  Maxime and Orio,  Aitor and Bordignon,  Anna and Bury,  Florian and Dorigo,  Tommaso and Giammanco,  Andrea and Heikal,  Mariam and Kieseler,  Jan and Lamparth,  Max and Martínez Ruíz del Árbol,  Pablo and Nardi,  Federico and Vischia,  Pietro and Zaraket,  Haitham},
  year = {2024},
  month = jul,
  pages = {035002}
}

@article{Tian2025,
  title = {Materials design with target-oriented Bayesian optimization},
  volume = {11},
  ISSN = {2057-3960},
  url = {http://dx.doi.org/10.1038/s41524-025-01704-4},
  DOI = {10.1038/s41524-025-01704-4},
  number = {1},
  journal = {npj Computational Materials},
  publisher = {Springer Science and Business Media LLC},
  author = {Tian,  Yuan and Li,  Tongtong and Pang,  Jianbo and Zhou,  Yumei and Xue,  Dezhen and Ding,  Xiangdong and Lookman,  Turab},
  year = {2025},
  month = jul 
}

@article{Schmidt2019,
  title = {Recent advances and applications of machine learning in solid-state materials science},
  volume = {5},
  ISSN = {2057-3960},
  url = {http://dx.doi.org/10.1038/s41524-019-0221-0},
  DOI = {10.1038/s41524-019-0221-0},
  number = {1},
  journal = {npj Computational Materials},
  publisher = {Springer Science and Business Media LLC},
  author = {Schmidt,  Jonathan and Marques,  Mário R. G. and Botti,  Silvana and Marques,  Miguel A. L.},
  year = {2019},
  month = aug 
}

@article{Regenwetter2022,
  title = {Deep Generative Models in Engineering Design: A Review},
  volume = {144},
  ISSN = {1528-9001},
  url = {http://dx.doi.org/10.1115/1.4053859},
  DOI = {10.1115/1.4053859},
  number = {7},
  journal = {Journal of Mechanical Design},
  publisher = {ASME International},
  author = {Regenwetter,  Lyle and Nobari,  Amin Heyrani and Ahmed,  Faez},
  year = {2022},
  month = mar 
}

@article{Olivecrona2017,
  title = {Molecular de-novo design through deep reinforcement learning},
  volume = {9},
  ISSN = {1758-2946},
  url = {http://dx.doi.org/10.1186/s13321-017-0235-x},
  DOI = {10.1186/s13321-017-0235-x},
  number = {1},
  journal = {Journal of Cheminformatics},
  publisher = {Springer Science and Business Media LLC},
  author = {Olivecrona,  Marcus and Blaschke,  Thomas and Engkvist,  Ola and Chen,  Hongming},
  year = {2017},
  month = sep 
}

@article{popova2018deep,
  title = {Deep reinforcement learning for de novo drug design},
  volume = {4},
  ISSN = {2375-2548},
  url = {http://dx.doi.org/10.1126/sciadv.aap7885},
  DOI = {10.1126/sciadv.aap7885},
  number = {7},
  journal = {Science Advances},
  publisher = {American Association for the Advancement of Science (AAAS)},
  author = {Popova,  Mariya and Isayev,  Olexandr and Tropsha,  Alexander},
  year = {2018},
  month = jul 
}

@article{Akmete2017,
  title = {The active muon shield in the SHiP experiment},
  volume = {12},
  ISSN = {1748-0221},
  url = {http://dx.doi.org/10.1088/1748-0221/12/05/P05011},
  DOI = {10.1088/1748-0221/12/05/p05011},
  number = {05},
  journal = {Journal of Instrumentation},
  publisher = {IOP Publishing},
  author = {Akmete,  A. and Alexandrov,  A. and Anokhina,  A. and Aoki,  S. and Atkin,  E. and Azorskiy,  N. and Back,  J.J. and Bagulya,  A. and Baranov,  A. and Barker,  G.J. and Bay,  A. and Bayliss,  V. and Bencivenni,  G. and Berdnikov,  A.Y. and Berdnikov,  Y.A. and Bertani,  M. and Betancourt,  C. and Bezshyiko,  I. and Bezshyyko,  O. and Bick,  D. and Bieschke,  S. and Blanco,  A. and Boehm,  J. and Bogomilov,  M. and Bondarenko,  K. and Bonivento,  W.M. and Boyarsky,  A. and Brenner,  R. and Breton,  D. and Brundler,  R. and Bruschi,  M. and B\"{u}scher,  V. and Buonaura,  A. and Buontempo,  S. and Cadeddu,  S. and Calcaterra,  A. and Campanelli,  M. and Chauveau,  J. and Chepurnov,  A. and Chernyavsky,  M. and Choi,  K.-Y. and Chumakov,  A. and Ciambrone,  P. and Dallavalle,  G.M. and D’Ambrosio,  N. and D’Appollonio,  G. and Lellis,  G. De and Roeck,  A. De and Serio,  M. De and Dedenko,  L. and Crescenzo,  A. Di and Marco,  N. Di and Dib,  C. and Dijkstra,  H. and Dmitrenko,  V. and Domenici,  D. and Donskov,  S. and Dubreuil,  A. and Ebert,  J. and Enik,  T. and Etenko,  A. and Fabbri,  F. and Fabbri,  L. and Fedin,  O. and Fedorova,  G. and Felici,  G. and Ferro-Luzzi,  M. and Fini,  R.A. and Fonte,  P. and Franco,  C. and Fukuda,  T. and Galati,  G. and Gavrilov,  G. and Gerlach,  S. and Golinka-Bezshyyko,  L. and Golubkov,  D. and Golutvin,  A. and Gorbunov,  D. and Gorbunov,  S. and Gorkavenko,  V. and Gornushkin,  Y. and Gorshenkov,  M. and Grachev,  V. and Graverini,  E. and Grichine,  V. and Guler,  A. M. and Guz,  Yu. and Hagner,  C. and Hakobyan,  H. and Herwijnen,  E. van and Hollnagel,  A. and Hosseini,  B. and Hushchyn,  M. and Iaselli,  G. and Iuliano,  A. and Jacobsson,  R. and Jonker,  M. and Kadenko,  I. and Kamiscioglu,  C. and Kamiscioglu,  M. and Khabibullin,  M. and Khaustov,  G. and Khotyantsev,  A. and Kim,  S.H. and Kim,  V. and Kim,  Y.G. and Kitagawa,  N. and Ko,  J.-W. and Kodama,  K. and Kolesnikov,  A. and Kolev,  D.I. and Kolosov,  V. and Komatsu,  M. and Konovalova,  N. and Korkmaz,  M.A. and Korol,  I. and Korol’ko,  I. and Korzenev,  A. and Kovalenko,  S. and Krasilnikova,  I. and Krivova,  K. and Kudenko,  Y. and Kurochka,  V. and Kuznetsova,  E. and Lacker,  H.M. and Lai,  A. and Lanfranchi,  G. and Lantwin,  O. and Lauria,  A. and Lebbolo,  H. and Lee,  K.Y. and Lévy,  J.-M. and Lopes,  L. and Lyubovitskij,  V. and Maalmi,  J. and Magnan,  A. and Maleev,  V. and Malinin,  A. and Mefodev,  A. and Mermod,  P. and Mikado,  S. and Mikhaylov,  Yu. and Milstead,  D.A. and Mineev,  O. and Montanari,  A. and Montesi,  M.C. and Morishima,  K. and Movchan,  S. and Naganawa,  N. and Nakamura,  M. and Nakano,  T. and Novikov,  A. and Obinyakov,  B. and Ogawa,  S. and Okateva,  N. and Owen,  P.H. and Paoloni,  A. and Park,  B.D. and Paparella,  L. and Pastore,  A. and Patel,  M. and Pereyma,  D. and Petrenko,  D. and Petridis,  K. and Podgrudkov,  D. and Poliakov,  V. and Polukhina,  N. and Prokudin,  M. and Prota,  A. and Rademakers,  A. and Ratnikov,  F. and Rawlings,  T. and Razeti,  M. and Redi,  F. and Ricciardi,  S. and Roganova,  T. and Rogozhnikov,  A. and Rokujo,  H. and Rosa,  G. and Rovelli,  T. and Ruchayskiy,  O. and Ruf,  T. and Samoylenko,  V. and Saputi,  A. and Sato,  O. and Savchenko,  E.S. and Schmidt-Parzefall,  W. and Serra,  N. and Shakin,  A. and Shaposhnikov,  M. and Shatalov,  P. and Shchedrina,  T. and Shchutska,  L. and Shevchenko,  V. and Shibuya,  H. and Shustov,  A. and Silverstein,  S.B. and Simone,  S. and Skorokhvatov,  M. and Smirnov,  S. and Sohn,  J.Y. and Sokolenko,  A. and Starkov,  N. and Storaci,  B. and Strolin,  P. and Takahashi,  S. and Timiryasov,  I. and Tioukov,  V. and Tosi,  N. and Treille,  D. and Tsenov,  R. and Ulin,  S. and Ustyuzhanin,  A. and Uteshev,  Z. and Vankova-Kirilova,  G. and Vannucci,  F. and Venkova,  P. and Vilchinski,  S. and Villa,  M. and Vlasik,  K. and Volkov,  A. and Voronkov,  R. and Wanke,  R. and Woo,  J.-K. and Wurm,  M. and Xella,  S. and Yilmaz,  D. and Yilmazer,  A.U. and Yoon,  C.S. and Zaytsev,  Yu.},
  year = {2017},
  month = may,
  pages = {P05011–P05011}
}

@article{Schmidt2025,
  title = {End-to-End Detector Optimization with Diffusion Models: A Case Study in Sampling Calorimeters},
  volume = {8},
  ISSN = {2571-712X},
  url = {http://dx.doi.org/10.3390/particles8020047},
  DOI = {10.3390/particles8020047},
  number = {2},
  journal = {Particles},
  publisher = {MDPI AG},
  author = {Schmidt,  Kylian and Kota,  Krishna Nikhil and Kieseler,  Jan and De Vita,  Andrea and Klute,  Markus and Abhishek and Aehle,  Max and Awais,  Muhammad and Breccia,  Alessandro and Carroccio,  Riccardo and Chen,  Long and Dorigo,  Tommaso and Gauger,  Nicolas R. and Lupi,  Enrico and Nardi,  Federico and Nguyen,  Xuan Tung and Sandin,  Fredrik and Willmore,  Joseph and Vischia,  Pietro},
  year = {2025},
  month = apr,
  pages = {47}
}

@inproceedings{attention,
author = {Vaswani, Ashish and Shazeer, Noam and Parmar, Niki and Uszkoreit, Jakob and Jones, Llion and Gomez, Aidan N. and Kaiser, \L{}ukasz and Polosukhin, Illia}, title = {Attention is all you need}, year = {2017}, isbn = {9781510860964}, publisher = {Curran Associates Inc.}, address = {Red Hook, NY, USA}, booktitle = {Proceedings of the 31st International Conference on Neural Information Processing Systems}, pages = {6000–6010}, numpages = {11}, location = {Long Beach, California, USA}, series = {NIPS'17} }

@online{openai_chatgpt_2022,
  author       = {{OpenAI}},
  title        = {Introducing {ChatGPT}},
  year         = {2022},
  month        = nov,
  day          = {30},
  url          = {https://openai.com/index/chatgpt/?utm_source=chatgpt.com},
  note         = {Accessed: 2025-12-18},
  organization = {OpenAI}
}

@inproceedings{reasoning2022_wei,
author = {Wei, Jason and Wang, Xuezhi and Schuurmans, Dale and Bosma, Maarten and Ichter, Brian and Xia, Fei and Chi, Ed H. and Le, Quoc V. and Zhou, Denny},
title = {Chain-of-thought prompting elicits reasoning in large language models},
year = {2022},
isbn = {9781713871088},
publisher = {Curran Associates Inc.},
address = {Red Hook, NY, USA},
booktitle = {Proceedings of the 36th International Conference on Neural Information Processing Systems},
articleno = {1800},
numpages = {14},
location = {New Orleans, LA, USA},
series = {NIPS '22}
}

@inproceedings{zero_shot_llms_2022,
author = {Kojima, Takeshi and Gu, Shixiang Shane and Reid, Machel and Matsuo, Yutaka and Iwasawa, Yusuke},
title = {Large language models are zero-shot reasoners},
year = {2022},
isbn = {9781713871088},
publisher = {Curran Associates Inc.},
address = {Red Hook, NY, USA},
booktitle = {Proceedings of the 36th International Conference on Neural Information Processing Systems},
articleno = {1613},
numpages = {15},
location = {New Orleans, LA, USA},
series = {NIPS '22}
}

@misc{wei_emergent_abilities_llms_2022,
  doi = {10.48550/ARXIV.2206.07682},
  url = {https://arxiv.org/abs/2206.07682},
  author = {Wei,  Jason and Tay,  Yi and Bommasani,  Rishi and Raffel,  Colin and Zoph,  Barret and Borgeaud,  Sebastian and Yogatama,  Dani and Bosma,  Maarten and Zhou,  Denny and Metzler,  Donald and Chi,  Ed H. and Hashimoto,  Tatsunori and Vinyals,  Oriol and Liang,  Percy and Dean,  Jeff and Fedus,  William},
  title = {Emergent Abilities of Large Language Models},
  publisher = {arXiv},
  year = {2022},
  copyright = {Creative Commons Attribution 4.0 International}
}

@article{Chiarello2024,
  title = {Generative large language models in engineering design: opportunities and challenges},
  volume = {4},
  ISSN = {2732-527X},
  url = {http://dx.doi.org/10.1017/pds.2024.198},
  DOI = {10.1017/pds.2024.198},
  journal = {Proceedings of the Design Society},
  publisher = {Cambridge University Press (CUP)},
  author = {Chiarello,  Filippo and Barandoni,  Simone and Majda Škec,  Marija and Fantoni,  Gualtiero},
  year = {2024},
  month = may,
  pages = {1959–1968}
}

@article{Gpfert2024,
  title = {Opportunities for large language models and discourse in engineering design},
  volume = {17},
  ISSN = {2666-5468},
  url = {http://dx.doi.org/10.1016/j.egyai.2024.100383},
  DOI = {10.1016/j.egyai.2024.100383},
  journal = {Energy and AI},
  publisher = {Elsevier BV},
  author = {G\"{o}pfert,  Jan and Weinand,  Jann M. and Kuckertz,  Patrick and Stolten,  Detlef},
  year = {2024},
  month = sep,
  pages = {100383}
}

@inproceedings{Lu2025,
  series = {ISPD ’25},
  title = {LEGO-Size: LLM-Enhanced GPU-Optimized Signoff-Accurate Differentiable VLSI Gate Sizing in Advanced Nodes},
  url = {http://dx.doi.org/10.1145/3698364.3705351},
  DOI = {10.1145/3698364.3705351},
  booktitle = {Proceedings of the 2025 International Symposium on Physical Design},
  publisher = {ACM},
  author = {Lu,  Yi-Chen and Kunal,  Kishor and Pradipta,  Geraldo and Liang,  Rongjian and Gandikota,  Ravikishore and Ren,  Haoxing},
  year = {2025},
  month = mar,
  pages = {152–162},
  collection = {ISPD ’25}
}

@article{Li2024,
  title = {LLM4CAD: Multimodal Large Language Models for Three-Dimensional Computer-Aided Design Generation},
  volume = {25},
  ISSN = {1944-7078},
  url = {http://dx.doi.org/10.1115/1.4067085},
  DOI = {10.1115/1.4067085},
  number = {2},
  journal = {Journal of Computing and Information Science in Engineering},
  publisher = {ASME International},
  author = {Li,  Xingang and Sun,  Yuewan and Sha,  Zhenghui},
  year = {2024},
  month = dec 
}

@article{Zheng2025,
  title = {Large language models for drug discovery and development},
  volume = {6},
  ISSN = {2666-3899},
  url = {http://dx.doi.org/10.1016/j.patter.2025.101346},
  DOI = {10.1016/j.patter.2025.101346},
  number = {10},
  journal = {Patterns},
  publisher = {Elsevier BV},
  author = {Zheng,  Yizhen and Koh,  Huan Yee and Ju,  Jiaxin and Yang,  Madeleine and May,  Lauren T. and Webb,  Geoffrey I. and Li,  Li and Pan,  Shirui and Church,  George},
  year = {2025},
  month = oct,
  pages = {101346}
}

@inproceedings{liu_llms_for_bayesian_2024,
 author = {Liu, Tennison and Astorga, Nicol\'{a}s and Seedat, Nabeel and van der Schaar, Mihaela},
 booktitle = {International Conference on Representation Learning},
 editor = {B. Kim and Y. Yue and S. Chaudhuri and K. Fragkiadaki and M. Khan and Y. Sun},
 pages = {31252--31284},
 title = {Large Language Models to Enhance Bayesian Optimization},
 url = {https://proceedings.iclr.cc/paper_files/paper/2024/file/84b8d9fcb4e262fcd429544697e1e720-Paper-Conference.pdf},
 volume = {2024},
 year = {2024}
}

@inproceedings{austin_llms_for_bayesian_2024,
author = {Austin, David and Korikov, Anton and Toroghi, Armin and Sanner, Scott},
title = {Bayesian Optimization with LLM-Based Acquisition Functions for Natural Language Preference Elicitation},
year = {2024},
isbn = {9798400705052},
publisher = {Association for Computing Machinery},
address = {New York, NY, USA},
url = {https://doi.org/10.1145/3640457.3688142},
doi = {10.1145/3640457.3688142},
booktitle = {Proceedings of the 18th ACM Conference on Recommender Systems},
pages = {74–83},
numpages = {10},
location = {Bari, Italy},
series = {RecSys '24}
}

@article{jin2025_wiseEDA,
title = {WiseEDA: LLMs in RF Circuit Design},
journal = {Microelectronics Journal},
volume = {158},
pages = {106607},
year = {2025},
issn = {1879-2391},
doi = {https://doi.org/10.1016/j.mejo.2025.106607},
url = {https://www.sciencedirect.com/science/article/pii/S1879239125000566},
author = {Hangjiang Jin and Junchao Wang and Junjie Sheng and Yifan Wu and Jiayu Chen and Yaqi Wang and Jun Liu}
}

@article{powell2009bobyqa,
  title={The BOBYQA algorithm for bound constrained optimization without derivatives},
  author={Powell, Michael JD and others},
  journal={Cambridge NA Report NA2009/06, University of Cambridge, Cambridge},
  volume={26},
  pages={26--46},
  year={2009}
}

@article{Cisbani2020,
  title = {AI-optimized detector design for the future Electron-Ion Collider: the dual-radiator RICH case},
  volume = {15},
  ISSN = {1748-0221},
  url = {http://dx.doi.org/10.1088/1748-0221/15/05/P05009},
  DOI = {10.1088/1748-0221/15/05/p05009},
  number = {05},
  journal = {Journal of Instrumentation},
  publisher = {IOP Publishing},
  author = {Cisbani,  E. and Dotto,  A. Del and Fanelli,  C. and Williams,  M. and Alfred,  M. and Barbosa,  F. and Barion,  L. and Berdnikov,  V. and Brooks,  W. and Cao,  T. and Contalbrigo,  M. and Danagoulian,  S. and Datta,  A. and Demarteau,  M. and Denisov,  A. and Diefenthaler,  M. and Durum,  A. and Fields,  D. and Furletova,  Y. and Gleason,  C. and Grosse-Perdekamp,  M. and Hattawy,  M. and He,  X. and Hecke,  H. van and Higinbotham,  D. and Horn,  T. and Hyde,  C. and Ilieva,  Y. and Kalicy,  G. and Kebede,  A. and Kim,  B. and Liu,  M. and McKisson,  J. and Mendez,  R. and Nadel-Turonski,  P. and Pegg,  I. and Romanov,  D. and Sarsour,  M. and Silva,  C.L. da and Stevens,  J. and Sun,  X. and Syed,  S. and Towell,  R. and Xie,  J. and Zhao,  Z.W. and Zihlmann,  B. and Zorn,  C.},
  year = {2020},
  month = may,
  pages = {P05009–P05009}
}

\appendix

\section{Prompts}
\label{app:prompts}
\subsection{Calorimeter}

\begin{Verbatim}[breaklines=true,breakanywhere=true]
You optimize a sampling calorimeter layout. Return ONLY a pure JSON object with keys 'z' and 't' (no prose, no code fences). Constraints: - Units in mm. 1 <= z[i] and z[i]+t[i] <= 1500.00 - t[i] in [0.12, 0.15, 0.20] (discrete set) - Total sum(t) <= 6.00 - No overlaps: for sorted z, ensure z[i]+t[i] <= z[i+1] - z and t must have the same length (number of sensors). - you must output real numbers: example of output to avoid is 10.012.512, be careful of dots and commas. - Use nearly the entire thickness budget: target sum(t) in [5.9, 6.0] mm; proposals with sum(t) < 5.8 mm are likely to be rejected. - Do NOT space layers uniformly. Avoid placing most layers only in the first ~20 mm; explore a broader z-span (e.g., tens to hundreds of mm) as appropriate. - Allow mild jitter (small random offsets) and local clustering in promising regions, while keeping all constraints. Targets (\%): em50->8, em100->5, had50->25, had100->18. Example: {"z":[10.0, 50.0], "t":[0.20, 0.12]} If you cannot comply, output exactly: {"z":[], "t":[]}.
\end{Verbatim}

For the TR optimization we add: 
\begin{Verbatim}[breaklines=true,breakanywhere=true]
After you propose a design, a dedicated local optimizer (NLOpt BOBYQA) will automatically refine while keeping t fixed, enforcing the hard geometry constraints.
\end{Verbatim}

\subsection{Spectrometer}

\begin{Verbatim}[breaklines=true,breakanywhere=true]
You optimize the layout of a magnetic spectrometer with discrete tracking stations. Return ONLY a pure JSON object with keys 'z' and 'g' (no prose, no code fences). Meaning of fields: - z[i]: position of tracking station i along the beamline, in meters. - g[i]: per-station granularity (number of bins per side in the tracker plane). Constraints: - 0 <= z[i] <= 20.0. - At least 3 stations must be before z=10 and at least 3 after z=10 for good performance. - g[i] must be integers from 500 to 1000 inclusive, in steps of 50. - The total pixel budget sum(g[i]\^2) must not exceed 6000000. - You must output valid real numbers; avoid malformed tokens like '10.012.512'. Objective: - Designs that use too few pixels (e.g. sum g[i]\^2 < 5.5e6) are typically suboptimal; prefer using most of the available budget while respecting all constraints. - Maximize reconstruction performance: high efficiency and low momentum resolution at 10 and 100 GeV. - Designs with very poor performance (few stations or all on one side) will be heavily penalized. Output format example: {"z":[0.8, 2.5, 4.5, 7.5, 10.5, 12.5], "g":[850, 800, 750, 700, 650, 600]} If you cannot propose a valid design, output exactly: {"z":[], "g":[]}.
\end{Verbatim}

For the TR optimization we add: 
\begin{Verbatim}[breaklines=true,breakanywhere=true]
After you propose a design, a dedicated local optimizer (NLOpt BOBYQA) will automatically refine the station positions z while keeping g fixed, enforcing the hard geometry constraints.
\end{Verbatim}

\section{Design Iteration Algorithm}
\label{app:algorithm}
% Appendix: LLM-based proposal-and-evaluate search
% Requires: \usepackage[ruled,vlined,linesnumbered]{algorithm2e}

\begin{algorithm}[H]
\caption{LLM-based proposal--evaluate search with optional $z$-only TR refinement}
\label{alg:llm_search}
\DontPrintSemicolon

\textbf{Inputs:} design specification $\mathcal{P}$, context window $L$, reserve $R$, reply budget $F$, iterations $I$, decoding params $\tau$,\\
\texttt{use\_TR} (boolean), TR budget $I_{\mathrm{TR}}$ (max objective evaluations).\\
\textbf{State:} accepted designs $\mathcal{A}$, best reward $r^\star$.\\

\For{$i \gets 1$ \KwTo $I$}{
    Build base prompt from $\mathcal{P}$\;
    Pack accepted designs into prompt under token budget $L - R - F$\;

    $\mathbf{x}_{\mathrm{raw}} \leftarrow \textsc{Propose}(\text{prompt}, \tau)$\;
    $\mathbf{x}_{0} \leftarrow \textsc{ProjectToFeasible}(\mathbf{x}_{\mathrm{raw}}, \mathcal{P})$\;
    $(\mathrm{metrics}_{0}, r_{0}) \leftarrow \textsc{Evaluate}(\mathbf{x}_{0})$\;

    $\mathbf{x}_{\mathrm{best}} \leftarrow \mathbf{x}_{0}$; \quad $\mathrm{metrics}_{\mathrm{best}} \leftarrow \mathrm{metrics}_{0}$; \quad $r_{\mathrm{best}} \leftarrow r_{0}$\;

    \If{\texttt{use\_TR}}{
        $(\mathbf{x}_{\mathrm{TR}}, \mathrm{metrics}_{\mathrm{TR}}, r_{\mathrm{TR}}) \leftarrow \textsc{TR\_Refine}(\mathbf{x}_{0}, \mathcal{P}, I_{\mathrm{TR}})$\;
        \If{$r_{\mathrm{TR}} > r_{\mathrm{best}}$}{
            $\mathbf{x}_{\mathrm{best}} \leftarrow \mathbf{x}_{\mathrm{TR}}$\;
            $\mathrm{metrics}_{\mathrm{best}} \leftarrow \mathrm{metrics}_{\mathrm{TR}}$\;
            $r_{\mathrm{best}} \leftarrow r_{\mathrm{TR}}$\;
        }
    }

    Log raw proposal $\mathbf{x}_{\mathrm{raw}}$, projected design $\mathbf{x}_{0}$, final evaluated design $\mathbf{x}_{\mathrm{best}}$, metrics, and rewards\;

    \If{$r_{\mathrm{best}} > r^\star$}{
        $r^\star \leftarrow r_{\mathrm{best}}$\;
        Add $\mathbf{x}_{\mathrm{best}}$ to $\mathcal{A}$\;
    }
}
\Return best-so-far design and reward\;
\end{algorithm}

\end{document}